\documentclass[aip,jap,amsmath,amssymb,reprint,groupedaddress]{revtex4-1}

\usepackage{graphicx}
\usepackage{array}[=2016-10-06]
\usepackage{dcolumn}
\usepackage{bm}
\usepackage[utf8]{inputenc}
\usepackage[T1]{fontenc}
\usepackage{mathptmx}

\usepackage{booktabs}
\usepackage{placeins}
\usepackage{etoolbox}
\usepackage{url}
\usepackage[hidelinks]{hyperref}

\makeatletter
\def\@email#1#2{%
 \endgroup
 \patchcmd{\titleblock@produce}
  {\frontmatter@RRAPformat}
  {\frontmatter@RRAPformat{\produce@RRAP{*#1\href{mailto:#2}{#2}}}\frontmatter@RRAPformat}
  {}{}
}%
\makeatother

\begin{document}

\title{Surrogate-Based Optimization of Tandem Twin-Rotor Wind Turbines Accounting for Wake Interactions}

\author{Yinong Tian}
\author{Shan Wang$^{*}$}
\email[Author to whom correspondence should be addressed: ]{shan.wang@centec.tecnico.ulisboa.pt}
\affiliation{Centre for Marine Technology and Ocean Engineering (CENTEC), Instituto Superior T\'ecnico, Universidade de Lisboa, Av. Rovisco Pais 1, 1049-001 Lisboa, Portugal}

\date{\today}

\begin{abstract}
Wake interaction in a tandem twin-rotor turbine couples power capture with thrust on the shared support. This study develops a time-resolved surrogate for upstream IEA Wind 22 MW and downstream NREL 5 MW rotors. It covers spacing ratios $0.01\leq L/D\leq0.20$ and tip-speed ratios $4\leq\lambda\leq11$. Vortex-method simulations of 160 configurations under turbulent inflow train per-time-node XGBoost models to predict 600 s power, thrust, and torque histories. A CNN-Attention-LSTM network extends these histories, and a genetic algorithm balances the system power and thrust coefficients. Ten-fold cross-validation gives median power-history $R^2$ values of 0.995 upstream and 0.958 downstream for unseen configurations, but the first stage cannot extrapolate in tip-speed ratio. With fixed pitch and prescribed rotor speed, the coefficients apply where both rotors reach the simulated tip-speed ratio below rated wind speed. Increasing spacing raises upstream power by up to 6\% and lowers downstream power by up to 62\%. System power changes by less than 4\% at fixed $\lambda$. The high-power end of the Pareto front lies near $\lambda=10$. At $L/D=0.20$, the system power coefficient is highest. This configuration has the lowest peak thrust among the high-power candidates.
\end{abstract}

\maketitle

\section{Introduction}\label{sec:introduction}

Global installed offshore wind capacity reached approximately 83.2 GW
by the end of 2024 \cite{ref1}. The International Energy Agency estimates
that the technical potential of offshore wind exceeds global
electricity demand \cite{ref2}. Deployment depends on the cost of
foundations, installation, electrical infrastructure, operation and
maintenance, and financing \cite{ref3,ref4}. Increasing power capture per
support structure is one way to address these costs. The design must
balance this additional power against the increase in aerodynamic loading.

Larger rotors reduce the number of foundations and installation
operations needed per unit of installed capacity \cite{ref4,ref5}.
Further increases in rotor size, however, face constraints on blade
mass, structural loads, manufacturing, transport, and installation
\cite{ref5}. Twin-rotor and multi-rotor systems distribute capacity
across several rotors on a shared structure \cite{ref6,ref7,ref8}.
Jamieson and Branney \cite{ref6} proposed a 20 MW-class multi-rotor
concept to limit the mass penalty of enlarging a single rotor.
Ghaisas et al. \cite{ref7} found with large-eddy simulation that tip and
turbine spacing affect wake recovery and power density in multi-rotor
wind farms. Ismaiel and Yoshida \cite{ref8} examined support
flexibility, tower dynamics, and aerodynamic interaction for a coplanar
system with two NREL 5 MW rotors. Infrastructure sharing has also been
explored through shared anchors and mooring lines in floating wind
farms \cite{ref9,ref10}. In the tandem twin-rotor turbine studied
here, a large upstream rotor and a smaller downstream rotor are mounted
on one tower (Figure~\ref{fig:1}a). The downstream rotor operates in the
upstream rotor's wake, so axial spacing directly controls the
interaction between the two rotors.

Coaxial twin-rotor studies show that this interaction depends on
spacing, rotor size, and operating point
\cite{Pacholczyk2020,Shen2024,Salah2024}. Rosenberg et al.
\cite{Rosenberg2014} placed a small auxiliary rotor upstream of a main
rotor to recover energy lost near the hub. Large-eddy simulations
of a twin-rotor turbine by the same group predicted 5--6\% higher
aerodynamic efficiency than a single rotor \cite{Moghadassian2016}.
Wind-tunnel tests of a design with a small upwind rotor measured a
7.2\% increase in power coefficient for the counter-rotating
arrangement \cite{Wang2018}. For a small counter-rotating turbine,
closer rotor spacing increased the power gain over a single rotor but
introduced rotor interactions that made torque and power fluctuate.
These fluctuations were not apparent beyond 0.5$D$ \cite{Pacholczyk2020}.
A blade-element-momentum model with two-way rotor interaction produced
power-coefficient maps with two peaks, which can lead a search to a
local maximum \cite{Shen2024}. Salah et al. \cite{Salah2024} found that
axial spacing had a negligible effect for equal rotor diameters and
examined the diameter ratio and downstream-rotor airfoil as further
variables. Li et al. \cite{Li2025} studied a 77 m upstream rotor and a
56 m downstream rotor. Total power increased by about 8\% as spacing
increased from 0.2$D$ to 2.0$D$, while downstream-rotor power decreased.
In tandem arrangements of separate turbines,
downstream power depends on both spacing and the upstream tip-speed
ratio \cite{Bayron2024}, and surge motion of floating multi-rotor
systems increases dynamic loads \cite{Arabgolarcheh2025}. Rotor
spacing, operating state, turbulence intensity, and platform motion
also affect twin-rotor wake recovery and loading
\cite{ref11,ref12}. Each configuration in these studies requires a
separate experiment or simulation, which limits the number of spacing
and tip-speed-ratio combinations that can be compared.

Vortex methods track wake evolution and mutual induction between
rotors \cite{ref13,ref14,Tian2026FVWreview}, and vortex-filament wake
models have been used to study the aerodynamic interference of
multi-unit turbines \cite{Liu2023multiunit}. They are therefore useful for generating
unsteady aerodynamic data, although repeated simulations can still be
costly in an optimization loop \cite{Tian2024annular}. Surrogate-based optimization
reduces this cost. Kriging and variable-fidelity surrogates based on
steady axisymmetric Reynolds-averaged Navier--Stokes (RANS) simulations
were used to optimize a twin-rotor turbine's secondary rotor
\cite{Thelen2018a,Thelen2018b}. They required less computational effort
than parametric sweeps or derivative-free search. Taghinezhad et al. \cite{Taghinezhad2021}
optimized a ducted twin-rotor turbine with a response-surface model.
Adema et al. \cite{Adema2026} used a wind-tunnel-validated
blade-element-momentum model to optimize the power performance of a
counter-rotating twin-rotor turbine. For pairs of vertical-axis rotors,
Fatahian et al. trained artificial neural networks on large-eddy
simulations and used multi-objective genetic algorithms to balance the
power coefficient against noise \cite{Fatahian2024,Fatahian2025}.
Rosenberg and Sharma \cite{Rosenberg2015} developed a prescribed-wake
vortex-line method for twin-rotor analysis and optimization, using time
averaging to handle rotors that turn at different speeds. These models
map design variables to scalar measures such as power coefficient,
energy production, or sound level. None of them predicts the thrust
history associated with a configuration.

Machine-learning models offer several ways to approximate nonlinear
wind-turbine responses. LSTM and CNN architectures have been used for
short-term wind-power forecasting and multi-turbine prediction
\cite{ref15,ref16,ref17}. Other studies have examined optimized LSTM
networks, Bayesian hyperparameter tuning, Gaussian process regression,
and support vector regression \cite{ref18,ref19,ref20}. Cascaded
networks \cite{ref24}, genetic feature selection \cite{ref25}, neural
generation of aggregate wind-power series \cite{ref26}, and gradient
boosting \cite{ref27} extend these forecasting tools. Wake-aware
prediction methods combine physical models with learned relationships
or use operating data to represent wake effects
\cite{ref21,ref22,ref23}. Related applications include
fatigue-load prediction under yaw control \cite{ref28}, wake-control
optimization for in-line turbines \cite{ref29},
acoustic-data-driven online power prediction \cite{ref30}, time-domain
motion prediction of floating platforms \cite{Liu2025XGBoost}, and
optimization of floating-turbine mooring systems under mooring-line
failure \cite{Li2025mooring}.
Sequential networks trained on aeroelastic simulations also map
inflow time series to damage-equivalent loads \cite{Haghi2024}. These
models learn temporal dependence for a turbine or wind farm whose
geometry and layout are fixed in the training data. They therefore do
not predict how the response of a tandem rotor pair changes with axial
spacing.

These two lines of work leave a gap for tandem rotors at close
spacing. Configuration surrogates return scalar averages, whereas close
spacing produces fluctuating torque and power \cite{Pacholczyk2020}.
Load measures such as the peak thrust are properties of the response
history, which a scalar surrogate does not provide. Time-series models
resolve that history but treat the configuration as fixed. To the
authors' knowledge, no previous study has predicted the complete power
and thrust histories of a tandem twin-rotor turbine as functions of
spacing and tip-speed ratio. No configuration search for such a turbine
has evaluated the power--thrust trade-off from these histories.
Table~\ref{tab:novelty} compares the present study with the most closely
related work.

The study examines an unequal pair of multi-megawatt reference
rotors, with an upstream IEA Wind 22 MW rotor (WT1) and a downstream
NREL 5 MW rotor (WT2) \cite{ref31,ref32}. The spacing and tip-speed-ratio
ranges are $0.01\leq L/D\leq0.20$ and $4\leq\lambda\leq11$.
Previous coaxial studies used small-scale rotors
\cite{Wang2018,Pacholczyk2020,Salah2024}, placed the smaller rotor
upstream \cite{Rosenberg2014,Moghadassian2016,Wang2018}, or considered
spacings of 0.2$D$ and above \cite{Li2025}.

The training data come from 160 unsteady
vortex-method simulations with turbulent inflow, in which the wake
filaments of both rotors evolve and induce velocities on each other
\cite{ref33}. Previous twin-rotor optimizations relied on steady RANS,
blade-element-momentum, or time-averaged prescribed-wake models, or
reduced large-eddy simulations to time averages
\cite{Thelen2018a,Adema2026,Rosenberg2015,Fatahian2024}.

A two-stage surrogate predicts response
histories from the configuration variables. XGBoost maps tip-speed
ratio and spacing ratio to power, thrust, and torque at each time node
of the first 600 s. A CNN-Attention-LSTM network then extends these
histories by a further 600 s, using vortex-method wake descriptors for
the downstream rotor. The configuration is a model input, unlike in
forecasting models \cite{ref15,ref16,ref17}, and the output is a time
history, unlike in scalar configuration surrogates
\cite{Thelen2018a,Taghinezhad2021,Fatahian2024}.

A multi-objective genetic algorithm
(GA) searches the first-stage surrogate for configurations that trade the
time-mean system power coefficient against the system thrust coefficient. The peak thrust of the same histories
then separates candidates of nearly equal power. Searches at fixed
tip-speed ratio and at fixed spacing, together with the separate power
of each rotor, show how and why the preferred spacing changes with
tip-speed ratio.

\begin{table*}[!t]
\caption{Comparison of the present study with closely related twin-rotor, tandem-rotor, and surrogate-modeling studies.}
\label{tab:novelty}
\centering
\small\setlength{\tabcolsep}{4pt}\renewcommand{\arraystretch}{1.10}
\begin{tabular}{>{\raggedright\arraybackslash}p{2.6cm}>{\raggedright\arraybackslash}p{3.5cm}>{\raggedright\arraybackslash}p{3.3cm}>{\raggedright\arraybackslash}p{3.1cm}>{\raggedright\arraybackslash}p{3.6cm}}
\toprule
\textbf{Study} & \textbf{Rotor configuration} & \textbf{Data source} & \textbf{Model} & \textbf{Predicted output} \\
\midrule
Thelen et al. \cite{Thelen2018a,Thelen2018b} & Coaxial twin-rotor turbine with secondary-rotor design optimization & Steady axisymmetric RANS & Kriging and variable-fidelity surrogates & Steady rotor performance \\
Adema et al. \cite{Adema2026} & Counter-rotating twin-rotor turbine, 1.6 m rotors & Blade-element momentum, wind-tunnel validated & Direct optimization & Power coefficient, energy production \\
Rosenberg and Sharma \cite{Rosenberg2015} & Twin-rotor turbine, small rotor upstream & Prescribed-wake vortex line & Time-averaged vortex model & Time-averaged loading and power \\
Fatahian et al. \cite{Fatahian2024,Fatahian2025} & Paired vertical-axis rotors & Large-eddy simulation & Neural network with multi-objective GA & Mean power coefficient and noise \\
Li et al. \cite{Li2025} & Tandem 77 m and 56 m rotors, 0.2$D$--2.0$D$ & Unsteady CFD & Parametric sweep & Rotor power and flow field \\
Haghi and Crawford \cite{Haghi2024} & Single NREL 5 MW rotor & Aeroelastic simulation & Temporal convolutional network & Damage-equivalent load \\
Present study & Tandem IEA 22 MW and NREL 5 MW rotors, $0.01\leq L/D\leq0.20$ & Unsteady vortex method with turbulent inflow & XGBoost and CNN-Attention-LSTM with GA & Power, thrust, and torque histories; mean power and thrust coefficients, peak thrust \\
\bottomrule
\end{tabular}
\end{table*}

Section~\ref{sec:methodology} describes the vortex model and its
validation, the two prediction stages, and the optimization procedure.
Section~\ref{sec:verification-and-discussion} compares the temporal
model with simpler alternatives, evaluates the two-stage predictions
within and beyond the sampled range, analyses the selected
configurations, and discusses computational cost and limitations.
Section~\ref{sec:conclusion} summarizes the findings.

\section{Model and Method}\label{sec:methodology}

\subsection{Twin-rotor model and the vortex method}\label{subsec:twin-rotor-model-and-the-vortex-method}

The aerodynamic model resolves the wake evolution of both rotors and
their mutual induction. Simulations use the unsteady lifting-line
vortex-lattice method implemented in MRUT, a filament-based free-vortex-wake
formulation \cite{Tian2026FVWreview} of the type used in earlier
multi-rotor and twin-rotor wake studies by the authors
\cite{Liu2023multiunit,ref11}. The software and related
twin-rotor power and wake data are archived in
References~\cite{ref33,ref34}.

\begin{figure*}[!t]
\centering
\includegraphics[width=0.94\linewidth,height=0.57\textheight,keepaspectratio]{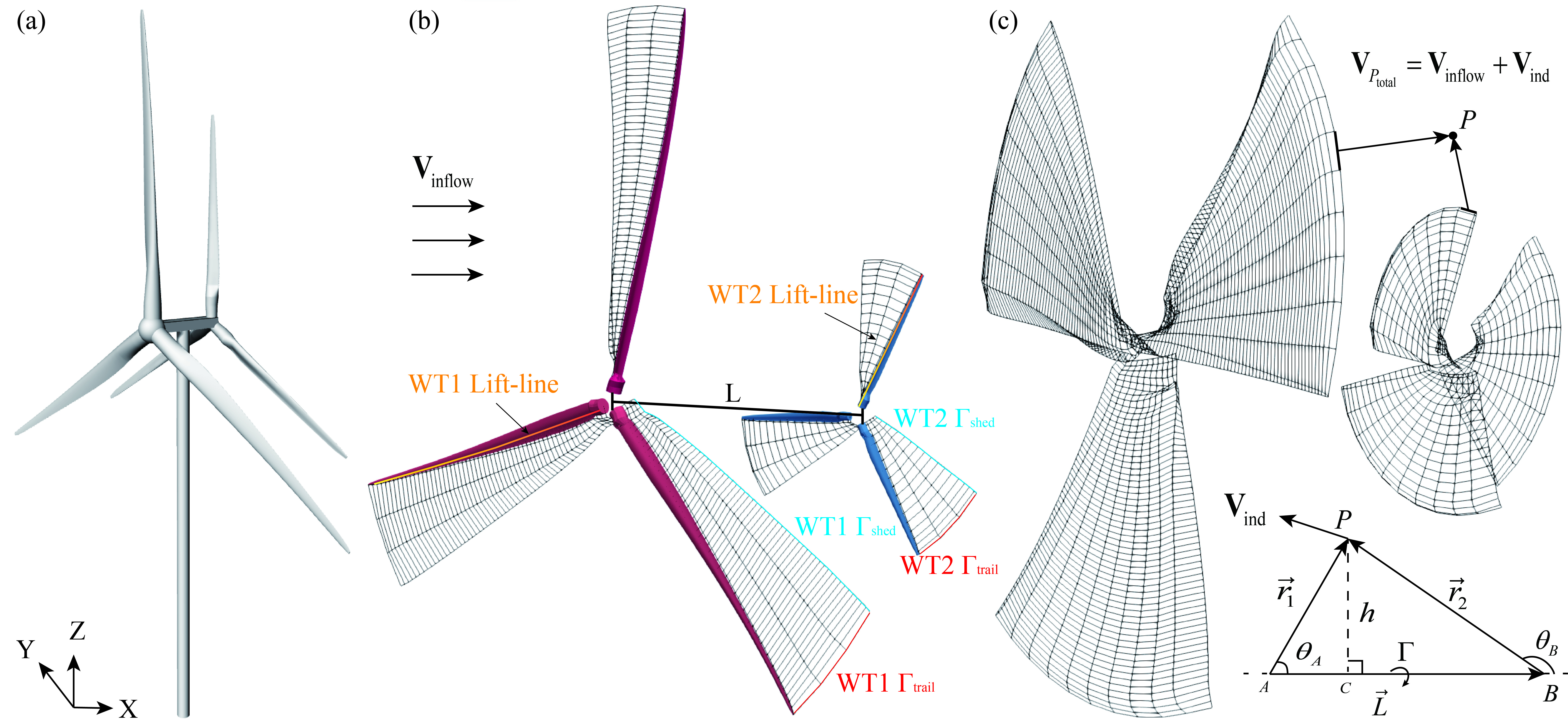}
\caption{Twin-rotor wind turbine simulation model and vortex method.}
\label{fig:1}
\end{figure*}

Figure~\ref{fig:1}a shows the tandem twin-rotor wind turbine.
WT1 is based on the IEA Wind 22 MW offshore reference turbine
\cite{ref31}, and WT2 on the NREL 5 MW offshore reference turbine
\cite{ref32}. Figure~\ref{fig:1}b shows the lifting-line distribution.
Figure~\ref{fig:1}c illustrates mutual wake induction between the vortex
filaments of both rotors over 10 time steps. The velocity at any point
is the sum of the freestream velocity and the velocities induced by
each rotor's bound, shed, and trailing vortices.

The lower-right schematic shows the velocity induced at point P
by a vortex-filament segment. It is computed using the Biot-Savart law:

\begin{equation}
\mathbf{V}_{\mathrm{induced},P}=
\sum_{m=\mathrm{WT1}}^{\mathrm{WT2}}\sum_{k\in W_m}
K_v\frac{\Gamma_k}{4\pi}
\frac{(r_1+r_2)(\mathbf{r}_1\times\mathbf{r}_2)}
{r_1r_2\left(r_1r_2+\mathbf{r}_1\cdot\mathbf{r}_2\right)} .
\label{eq:biot-savart}
\end{equation}

Here, $\mathbf{r}_1$ and $\mathbf{r}_2$ are the vectors from the
segment endpoints A and B to P, and $r_1$ and $r_2$ are their
magnitudes. $W_m$ denotes the vortex filaments associated with rotor
$m$, and $\Gamma_k$ is the circulation of filament $k$. The Vatistas
factor $K_v$ regularizes the induced velocity using the vortex-core
radius $r_c$ \cite{ref35}:

\begin{equation}
K_v=\frac{(d/r_c)^2}{\sqrt{1+(d/r_c)^4}},\qquad
d=\begin{cases}
r_1, & \cos\beta_1<0,\\
r_2, & \cos\beta_2>0,\\
h,   & \text{otherwise}.
\end{cases}
\label{eq:vatistas}
\end{equation}

\begin{equation}
\left\{
\begin{aligned}
\mathbf{r}_0 &= \mathbf{r}_1-\mathbf{r}_2,\\
\cos\beta_1 &= \frac{\mathbf{r}_1\cdot\mathbf{r}_0}{r_1r_0},\\
\cos\beta_2 &= \frac{\mathbf{r}_2\cdot\mathbf{r}_0}{r_2r_0},\\
h &= \frac{\left\lvert\mathbf{r}_1\times\mathbf{r}_2\right\rvert}
{\left\lvert\mathbf{L}\right\rvert}.
\end{aligned}
\right.
\label{eq:vortex-geometry}
\end{equation}

In Equation~\eqref{eq:vortex-geometry}, $\mathbf{L}$ is the vortex-segment
vector and $h$ is the perpendicular distance from P to its supporting
line. Vortex-core growth and filament stretching \cite{Tian2026FVWreview} are represented by

\begin{equation}
\left\{
\begin{aligned}
\varepsilon &= \frac{\Delta l}{l},\\
r_c(\zeta) &= \sqrt{r_{c0}^{2}+
\frac{4\alpha_L\delta\nu\zeta}{(1+\varepsilon)\omega}}.
\end{aligned}
\right.
\label{eq:core-growth}
\end{equation}

Here, $l$ is the filament length before stretching, $\Delta l$ is its
change in length, and $r_{c0}$ is the initial core radius.
$\zeta$ denotes wake age in angular form and $\omega$ is rotor angular
velocity. For the Lamb-Oseen model, $\alpha_L=1.25643$. The
vortex-viscosity parameter $\delta$ ranges from 10 to 1000 depending on
rotor size \cite{ref36}. The kinematic viscosity used here is
$\nu=1.647\times10^{-5}$ m$^2$/s. The present simulations use
$\delta=1000$ and an initial core radius $r_{c0}$ equal to 10\% of the
local chord at the element midpoint. The Vatistas core model includes
core growth and filament stretching. Filaments are
convected with a first-order explicit Euler scheme and contribute to the
induction over a near-wake length of $5D$. The core
parameters set how quickly the wake vorticity diffuses. In a related
vortex-ring model \cite{Tian2024annular}, the authors examined core radii expressed as fractions
of blade length. The near-wake velocity profile changed markedly with
the initial core radius. Values of 3--20\% gave consistent profiles.
A value of 50\% smoothed out the tip- and root-vortex structure, whereas
values of 1\% or less produced overly sharp velocity minima. Because WT2 operates
in the near wake of WT1, these parameters affect the WT2 inflow and are a
source of model uncertainty in the downstream predictions.

The resultant velocity at the control point of each blade element is
calculated as follows:

\begin{equation}
\mathbf{V}_{\mathrm{total}}=\mathbf{V}_{\mathrm{inflow}}+
\mathbf{V}_{\mathrm{ind}}.
\label{eq:total-velocity}
\end{equation}

Here, $\mathbf{V}_{\mathrm{inflow}}$ is the three-dimensional freestream
velocity, and $\mathbf{V}_{\mathrm{ind}}$ is the velocity induced by the
bound, shed, and trailing filaments.

All configurations use the same turbulent inflow, generated with
TurbSim \cite{Jonkman2009} at a mean wind speed of 11.4 m/s and a
turbulence intensity of 10\%. The 600 s wind sequence is repeated when a
simulation is longer than one cycle. Each of the 160 grid simulations in
Section~\ref{subsec:multi-parameter-power-prediction-for-twin-rotor-wind-turbines}
covers a 60 s startup interval and one 600 s cycle. Records contain
4236 time nodes between 60 s and 660 s, at a constant interval of
0.1417 s. Each rotor's records include power, thrust, torque,
aerodynamic coefficients, instantaneous tip-speed ratio, and inflow
components. The verification case and the six
evaluation conditions in Section~\ref{sec:verification-and-discussion}
cover 1260 s, that is, the startup interval and two cycles.
Over the 600 s window, the mean axial inflow recorded at the rotors is
11.54 m/s.

The vortex-method simulations use fixed blade pitch and prescribed
rotor speed. No pitch or torque controller is modeled. Both
rotors share the nominal tip-speed ratio $\lambda$, defined with the
mean wind speed $U_\infty=11.4$ m/s. The rotor angular velocities are
then $\omega_m=2\lambda U_\infty/D_m$, where $D_1=284$ m and $D_2=126$ m
are the rotor diameters. For fixed pitch and fixed $\lambda$, the mean power
and thrust scale with $U_\infty^3$ and $U_\infty^2$. They are reported as
coefficients referred to $U_\infty$ and to the swept area of the
upstream rotor, which encloses the downstream rotor. The system power
and thrust coefficients are
\[
C_P=\frac{P_1+P_2}{\tfrac{\pi}{8}\rho U_\infty^3D_1^2},\qquad
C_T=\frac{T_1+T_2}{\tfrac{\pi}{8}\rho U_\infty^2D_1^2},
\]
with $\rho=1.225$ kg/m$^3$. The rotor coefficients $C_{P,m}$ and
$C_{T,m}$ are defined in the same way with the swept area of rotor $m$,
and $(D_2/D_1)^2=0.197$. Apart from Reynolds-number effects and
the wind-speed dependence of the turbulent inflow, the coefficients
apply wherever both rotors can run at the given $\lambda$ within their
rotor-speed ranges.

\begin{figure}[!htb]
\centering
\includegraphics[width=\columnwidth]{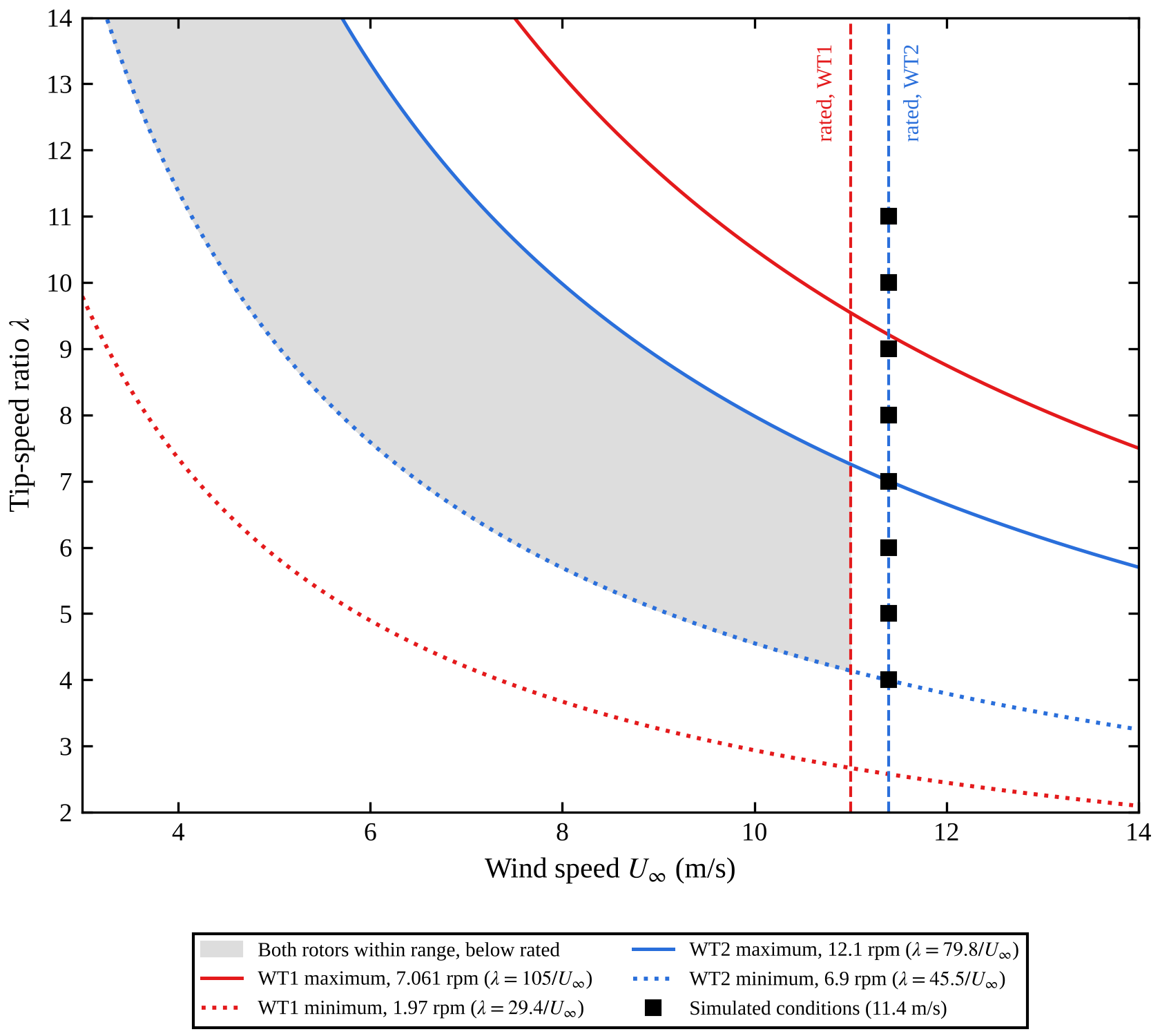}
\caption{Operating envelope of the two reference rotors for a shared tip-speed ratio.}
\label{fig:envelope}
\end{figure}

Figure~\ref{fig:envelope} shows this operating envelope. Its solid and
dotted curves give the tip-speed ratio at the maximum and minimum rotor
speeds of each turbine, the dashed lines mark the rated wind speeds, and
the squares mark the simulated conditions at 11.4 m/s. The shaded region
contains the wind speeds between cut-in and the WT1 rated wind speed at
which both rotors are within their rotor-speed ranges. The IEA Wind
22 MW turbine has a rated wind speed of 11.0 m/s and a maximum rotor
speed of 7.061 rpm, at a tip speed of 105 m/s \cite{ref31}. Its cut-in
wind speed is 3 m/s, and the minimum rotor speed of its controller is
1.97 rpm \cite{IEA22data}. The NREL 5 MW turbine has cut-in and rated
wind speeds of 3 and 11.4 m/s and cut-in and rated rotor speeds of 6.9
and 12.1 rpm \cite{ref32}. With a shared $\lambda$, both bounds are set
by WT2: $45.5/U_\infty\leq\lambda\leq79.8/U_\infty$, with $U_\infty$ in
m/s. Below the rated wind speed of WT1, $\lambda=4$ cannot be reached,
$\lambda=5$ requires 9.1--11.0 m/s, and $\lambda=10$ requires
4.6--8.0 m/s. The simulated wind speed of 11.4 m/s equals that of the
CFD validation case described below. At this wind speed, the
fixed-pitch power of WT1 exceeds its rated 22 MW for $\lambda\geq7$,
whereas the power of WT2 stays below its rated 5 MW (at most 4.0 MW).
Above rated wind speed, the controllers would pitch the blades to limit
the power. Dimensional values are given only as examples at wind speeds
inside the envelope.

MRUT was compared with a blade-resolved computational fluid
dynamics (CFD) model of the same tandem pair at $L/D=0.1$.
Star-CCM+ solved the unsteady Reynolds-averaged Navier--Stokes (RANS)
equations with the SST $k$--$\omega$ model \cite{Menter1994}. The
overset rotor meshes contained $2.4\times10^7$ cells. The time step was
0.0413 s, corresponding to $3^\circ$ of WT2 rotation. Both were selected
through grid- and time-step-independence tests. Both codes used a
uniform inflow of 11.4 m/s. Rotor speeds were the rated speeds of
7.06 rpm for WT1 and 12.1 rpm for WT2, which correspond to tip-speed
ratios of 9.2 and 7.0. This case is not a point of the simulation grid. The mean thrust, torque, and power
coefficients of the two codes differ by at most 0.5\%
(Table~\ref{tab:validation}). MRUT gives smoother load histories with
fewer high-frequency components than the RANS solution. Its wake
profiles agree more closely with the CFD profiles at $2D$ and $3D$
downstream than at $0.5D$ and $1D$. At blade level,
MRUT spanwise circulation and angle-of-attack distributions for the
NREL 5 MW rotor agree with QBlade \cite{ref14}. The authors also tested
a model-scale tandem pair with rotor diameters of 0.8 m and 0.6 m in a
wind tunnel. These blades were designed for the tunnel and are not
geometrically similar to the present rotors. Tests at $L/D=0.2$ and 0.4
showed the same qualitative trend. Larger spacing raised upstream-rotor
power and lowered the downstream-rotor contribution. At $L/D=0.4$ with
counter-rotating rotors, MRUT with a tower-shadow correction predicted
upstream-rotor thrust within $\pm4\%$ but overpredicted downstream-rotor
thrust. The latter discrepancy is attributed to underpredicted
turbulent dissipation in the upstream wake. The validation covers mean
loads of the present rotor pair at one spacing and spacing trends at
model scale. The turbulent load histories used for training have no
direct experimental reference.

\begin{table*}[!t]
\caption{Mean rotor coefficients of MRUT and blade-resolved RANS CFD for the tandem IEA Wind 22 MW and NREL 5 MW pair at $L/D=0.1$, uniform inflow of 11.4 m/s, and rotor speeds of 7.06 rpm (WT1) and 12.1 rpm (WT2).}
\label{tab:validation}
\centering
\small\setlength{\tabcolsep}{4pt}\renewcommand{\arraystretch}{1.10}
\begin{tabular}{lcccccc}
\toprule
 & \multicolumn{3}{c}{\textbf{WT1}} & \multicolumn{3}{c}{\textbf{WT2}} \\
\cmidrule(lr){2-4}\cmidrule(lr){5-7}
\textbf{Coefficient} & \textbf{MRUT} & \textbf{CFD} & \textbf{Difference} & \textbf{MRUT} & \textbf{CFD} & \textbf{Difference} \\
\midrule
$C_T$ & 0.8297 & 0.8314 & $-0.20\%$ & 0.5008 & 0.5033 & $-0.50\%$ \\
$C_Q$ & 0.0515 & 0.0516 & $-0.19\%$ & 0.0247 & 0.0248 & $-0.40\%$ \\
$C_P$ & 0.4744 & 0.4753 & $-0.19\%$ & 0.1802 & 0.1811 & $-0.50\%$ \\
\bottomrule
\end{tabular}
\end{table*}

\subsection{CNN-Attention-LSTM prediction model}\label{subsec:cnn-attention-lstm-prediction-model}

The temporal model combines one-dimensional convolution, attention
\cite{Bahdanau2015}, and LSTM \cite{Hochreiter1997} layers to
predict aerodynamic response histories. The CNN
extracts local temporal patterns, the attention layer weights the
resulting sequence, and the LSTM carries information between time
steps. These components are intended to represent the fluctuations
associated with turbulent inflow and rotor interaction.

\begin{figure*}[!t]
\centering
\includegraphics[width=0.94\linewidth,height=0.57\textheight,keepaspectratio]{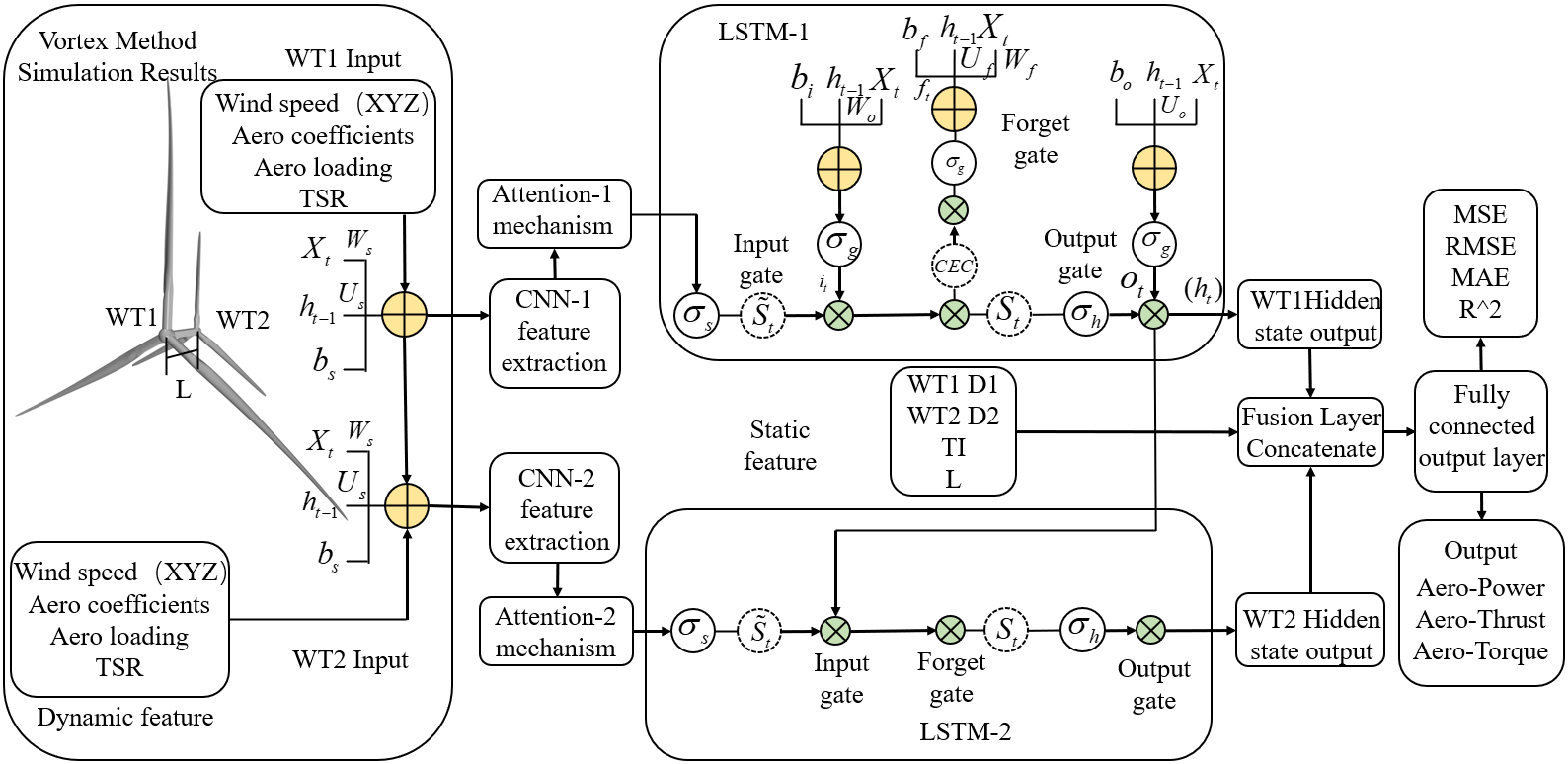}
\caption{Aerodynamic-response prediction workflow for the tandem twin-rotor wind turbine. In the sketch, $L$ is the distance between the rotor centres. The static feature labelled $L$ is the spacing ratio $L/D$.}
\label{fig:2}
\end{figure*}

Figure~\ref{fig:2} shows the two prediction branches. Dynamic features
include wind-speed components, aerodynamic coefficients, response
histories, and tip-speed ratio. The configuration features are the
rotor diameters $D_1$ and $D_2$, turbulence intensity $TI$, and spacing
ratio $L/D$. These features remain fixed within a simulation.

The WT1 branch processes its own dynamic features and maps the LSTM
hidden state to power, thrust, and torque. The WT2 branch also receives
WT1 dynamic features to represent upstream influence. Its LSTM hidden
state is concatenated with the configuration features and vortex-method
wake descriptors, such as effective velocity at WT2, before the output
layer. Thus, the two branches have the same temporal architecture but
different input compositions.

Equation~\eqref{eq:lstm-gates} gives the activation vectors of the input gate, output
gate, and forget gate in the LSTM model:

\begin{equation}
\left\{
\begin{aligned}
i_t &= \sigma_i(W_i x_t+U_i h_{t-1}+b_i),\\
o_t &= \sigma_o(W_o x_t+U_o h_{t-1}+b_o),\\
f_t &= \sigma_f(W_f x_t+U_f h_{t-1}+b_f).
\end{aligned}
\right.
\label{eq:lstm-gates}
\end{equation}

Here, $W$ and $U$ are input and recurrent weight matrices, $b$ denotes
biases, and the $\sigma$ terms denote activation functions. The input
at step $t$ is $x_t$, the preceding hidden state is $h_{t-1}$, and
$S_t$ denotes the cell state. The candidate cell state is

\begin{equation}
\widetilde{S}_t=\sigma_s(W_s x_t+U_s h_{t-1}+b_s).
\label{eq:candidate-state}
\end{equation}

The gates control how the candidate and previous cell states are
combined and how the updated state contributes to the hidden output:

\begin{equation}
S_t=f_t\odot S_{t-1}+i_t\odot\widetilde{S}_t.
\label{eq:cell-state}
\end{equation}

\begin{equation}
h_t=o_t\odot\sigma_h(S_t).
\label{eq:hidden-state}
\end{equation}

The symbol $\odot$ denotes elementwise multiplication. The input gate
controls new information, the forget gate controls retention of the
previous state, and the output gate controls the hidden output. The
WT1 branch is summarized by

\begin{equation}
\left\{
\begin{aligned}
x_t^{\mathrm{WT1}}&=[U_1(t),C_1(t),R_1(t),\lambda_1(t)],\\
E_t^{\mathrm{WT1}}&=\operatorname{CNN}(x_t^{\mathrm{WT1}}),\\
A_t^{\mathrm{WT1}}&=\operatorname{Attention}(E_t^{\mathrm{WT1}}),\\
h_t^{\mathrm{WT1}}&=\operatorname{LSTM}_1(A_t^{\mathrm{WT1}},h_{t-1}^{\mathrm{WT1}}),\\
\operatorname{Output}_{\mathrm{WT1}}(t)&=[\widehat{P}_1(t),\widehat{T}_1(t),\widehat{Q}_1(t)]\\
&=W_{\mathrm{out}}^{\mathrm{WT1}}h_t^{\mathrm{WT1}}+b_{\mathrm{out}}^{\mathrm{WT1}}.
\end{aligned}
\right.
\label{eq:wt1-branch}
\end{equation}

In Equation~\eqref{eq:wt1-branch}, $U_1$, $C_1$, and $R_1$ group
the wind-speed components, aerodynamic coefficients, and aerodynamic
responses, respectively. The coefficients are $C_P$, $C_Q$, and $C_T$,
and the responses are power, thrust, and torque. The instantaneous
tip-speed ratio is $\lambda_1$. The predicted power, thrust, and torque
are $\widehat{P}_1$, $\widehat{T}_1$, and $\widehat{Q}_1$. $E$ denotes the convolutional feature
maps and $A$ the attention-weighted sequence. The equations summarize
the network mapping, with $t$ indexing the processed sequence. The
corresponding WT2 branch is

\begin{equation}
\left\{
\begin{aligned}
\mathbf{F}_{D}^{\mathrm{WT1}}(t)&=[U_1(t),C_1(t),R_1(t),\lambda_1(t)],\\
\mathbf{F}_{D}^{\mathrm{WT2}}(t)&=[U_2(t),C_2(t),R_2(t),\lambda_2(t)],\\
\mathbf{F}_{S}&=[L/D,D_1,D_2,TI],\\
\mathbf{F}_{VM}(t)&=[U_{\mathrm{eff},2}(t),\Delta U_2(t),\Delta TI_2(t)],\\
x_t^{\mathrm{WT2}}&=[\mathbf{F}_{D}^{\mathrm{WT1}}(t),\mathbf{F}_{D}^{\mathrm{WT2}}(t)],\\
E_t^{\mathrm{WT2}}&=\operatorname{CNN}(x_t^{\mathrm{WT2}}),\\
A_t^{\mathrm{WT2}}&=\operatorname{Attention}(E_t^{\mathrm{WT2}}),\\
h_t^{\mathrm{WT2}}&=\operatorname{LSTM}_2(A_t^{\mathrm{WT2}},h_{t-1}^{\mathrm{WT2}}),\\
\mathbf{F}_{\mathrm{merged}}(t)&=\operatorname{Concat}(h_t^{\mathrm{WT2}},\mathbf{F}_{S},\mathbf{F}_{VM}(t)),\\
[\widehat{P}_2(t),\widehat{T}_2(t),\widehat{Q}_2(t)]&=W_{\mathrm{out}}^{\mathrm{WT2}}\mathbf{F}_{\mathrm{merged}}(t)+b_{\mathrm{out}}^{\mathrm{WT2}}.
\end{aligned}
\right.
\label{eq:wt2-branch}
\end{equation}

In Equation~\eqref{eq:wt2-branch}, $\mathbf{F}_{D}^{\mathrm{WT1}}$ and
$\mathbf{F}_{D}^{\mathrm{WT2}}$ contain the rotor-specific dynamic
features, while $\mathbf{F}_{S}$ contains the fixed configuration
features. The vortex-method descriptors are grouped in
$\mathbf{F}_{VM}$. They comprise the effective axial velocity at
WT2, $U_{\mathrm{eff},2}$, its freestream deficit, $\Delta U_2$, and the
local turbulence enhancement, $\Delta TI_2$. All are obtained from the
vortex-method induction. The output layer acts on the
concatenated vector $\mathbf{F}_{\mathrm{merged}}$. These grouped feature
definitions describe the branch architecture. The specific ten non-time
variables used in the input comparison are listed in
Section~\ref{subsec:lstm-model-verification}. The descriptors in
$\mathbf{F}_{VM}$ are model outputs and are not measured on an operating
turbine. In training and verification, they are taken from the
simulated histories (Section~\ref{subsec:lstm-model-verification}). When
the two-stage pipeline of
Section~\ref{subsec:multi-parameter-prediction-model-based-on-xgboost}
is applied to a new configuration, they are computed with the vortex
method. The configuration search of
Section~\ref{subsec:twin-rotor-configuration-parameter-optimization-model-based-on-ga}
uses only the first stage and needs no vortex-method input.

\subsection{Multi-parameter prediction model based on XGBoost}\label{subsec:multi-parameter-prediction-model-based-on-xgboost}

XGBoost \cite{Chen2016}, which has also been used to predict the
time-domain motion of floating platforms \cite{Liu2025XGBoost}, maps tip-speed ratio and spacing ratio to
the first-stage aerodynamic response. MRUT simulations provide the
training targets for WT1 and WT2 after the startup transient is
excluded. A separate regression model is fitted at each time node, and
the nodal predictions form a response history that can be passed to
CNN-Attention-LSTM. The inputs are $L/D$, $\lambda$, the rotor
index, and the engineered terms $\lambda L/D$ and $\lambda^2$.
Hyperparameters were chosen by randomized search with three-fold
cross-validation over the training configurations
(Table~\ref{tab:hyperparameters}). The models are trained on all 160
grid configurations and tested on six further MRUT simulations at
conditions that are not grid points
(Section~\ref{subsec:multi-parameter-power-prediction-for-twin-rotor-wind-turbines}).

For a given time node and response variable, the training data comprise
$n$ configuration vectors $x_i$ and their simulation responses $p_i$.
An ensemble of $K$ regression trees, $\{f_k\}_{k=1}^{K}$, minimizes

\begin{equation}
\operatorname{Obj}=\sum_{i=1}^{n}l(p_i,\widehat{p}_i)
+\sum_{k=1}^{K}\Omega(f_k).
\label{eq:xgb-objective}
\end{equation}

Here, $l$ is the loss, $\widehat{p}_i$ is the predicted response for
sample $i$, and $\Omega$ penalizes tree complexity. At boosting
iteration $t$, a second-order expansion around the current prediction
$\widehat{p}_i^{(t-1)}$ gives the following objective after terms
independent of the new tree are omitted:

\begin{equation}
\left\{
\begin{aligned}
\operatorname{Obj}^{(t)}&=\sum_{i=1}^{n}
\left[g_if_t(x_i)+\frac{1}{2}h_if_t^2(x_i)\right]
+\Omega(f_t),\\
g_i&=\partial_{\widehat{p}_i^{(t-1)}}l(p_i,\widehat{p}_i^{(t-1)}),\\
h_i&=\partial_{\widehat{p}_i^{(t-1)}}^2l(p_i,\widehat{p}_i^{(t-1)}).
\end{aligned}
\right.
\label{eq:xgb-taylor}
\end{equation}

The index $t$ in this derivation denotes a boosting iteration rather
than physical time. The quantities $g_i$ and $h_i$ are the first and
second derivatives of the loss with respect to the current prediction.
For a tree with $T$ leaves and leaf weights $w_j$, the regularization
term and the objective grouped by leaf are

\begin{equation}
\Omega(f_t)=\gamma T+\frac{1}{2}\lambda_r\sum_{j=1}^{T}w_j^2.
\label{eq:xgb-regularization}
\end{equation}

\begin{equation}
\begin{aligned}
\operatorname{Obj}^{(t)}
&=\sum_{i=1}^{n}\left[g_iw_{q(x_i)}+\frac{1}{2}h_iw_{q(x_i)}^2\right]
+\gamma T+\frac{1}{2}\lambda_r\sum_{j=1}^{T}w_j^2\\
&=\sum_{j=1}^{T}\left[\left(\sum_{i\in I_j}g_i\right)w_j
+\frac{1}{2}\left(\sum_{i\in I_j}h_i+\lambda_r\right)w_j^2\right]+\gamma T.
\end{aligned}
\label{eq:xgb-leaves}
\end{equation}

Here, $\gamma$ penalizes the number of leaves and $\lambda_r$ penalizes
their weights. The subscript distinguishes $\lambda_r$ from the
tip-speed ratio $\lambda$. The mapping $q(x_i)$ assigns configuration
$x_i$ to a leaf, and $I_j$ contains the samples assigned to leaf $j$.
For a fixed tree structure, substituting the minimizing leaf weights
gives

\begin{equation}
\left\{
\begin{aligned}
\operatorname{Obj}^{(t)}&=-\frac{1}{2}\sum_{j=1}^{T}
\frac{G_j^2}{H_j+\lambda_r}+\gamma T,\\
G_j&=\sum_{i\in I_j}g_i,\\
H_j&=\sum_{i\in I_j}h_i.
\end{aligned}
\right.
\label{eq:xgb-final}
\end{equation}

The quantities $G_j$ and $H_j$ are the sums of the loss derivatives
over leaf $j$.

\begin{figure*}[!t]
\centering
\includegraphics[width=0.90\linewidth,height=0.57\textheight,keepaspectratio]{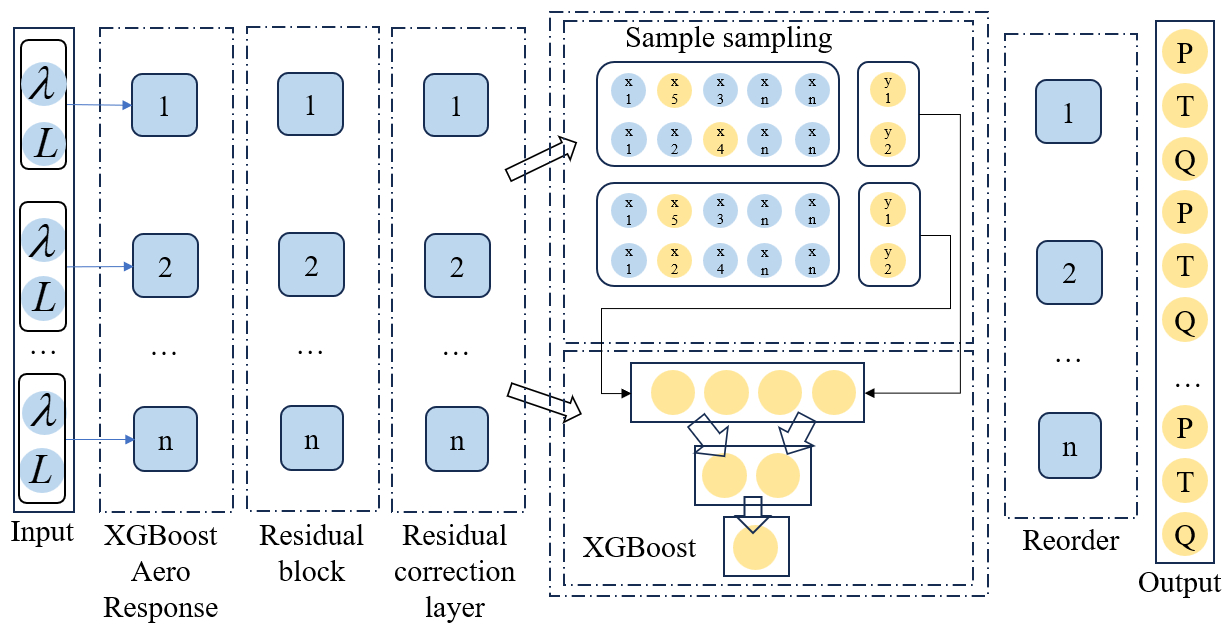}
\caption{XGBoost regression-tree training and response prediction. The inputs are the tip-speed ratio $\lambda$ and the spacing ratio $L/D$, labelled $L$ in the figure.}
\label{fig:3}
\end{figure*}

\begin{figure*}[!t]
\centering
\includegraphics[width=0.90\linewidth,height=0.57\textheight,keepaspectratio]{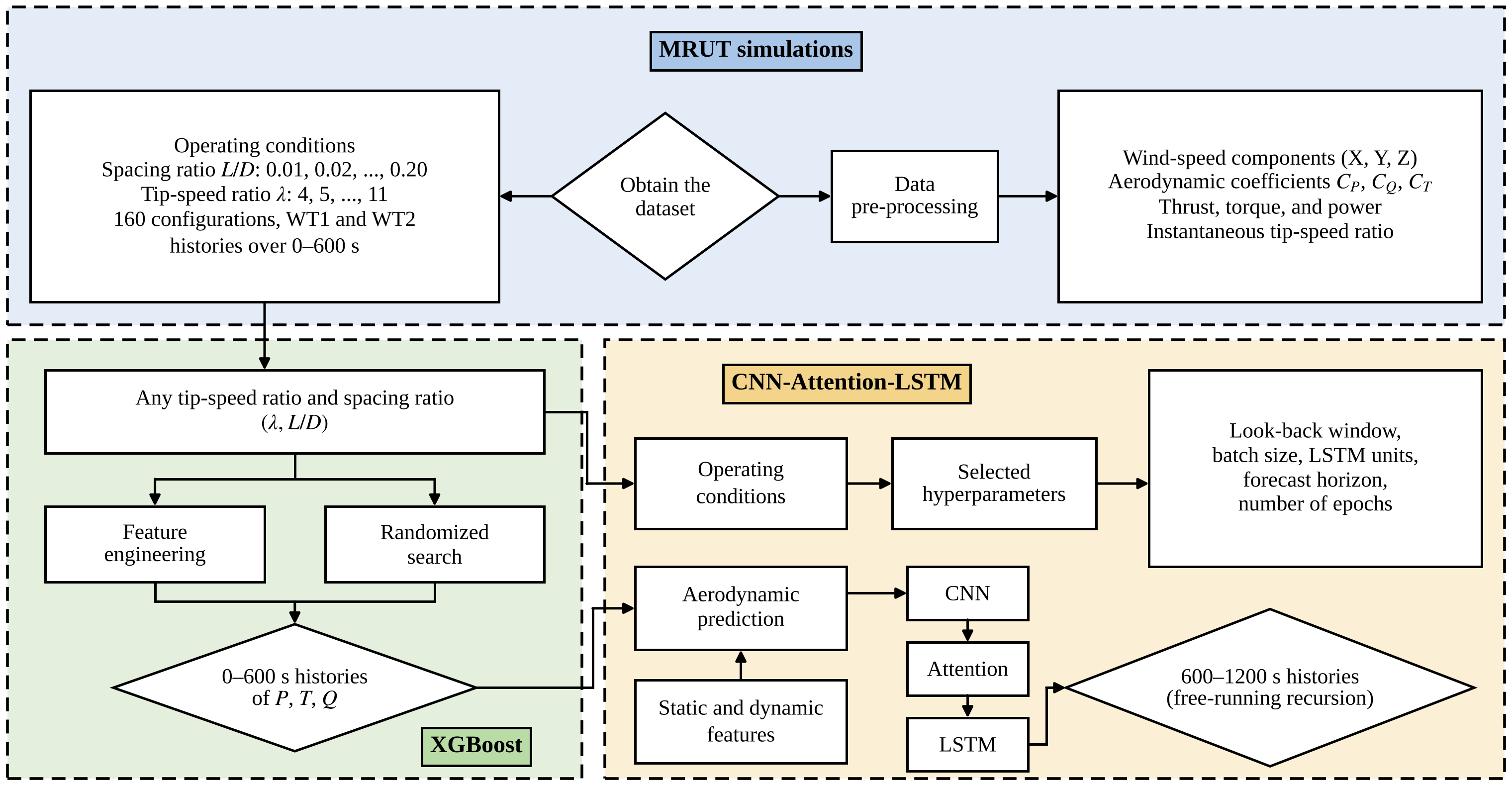}
\caption{Combined XGBoost and CNN-Attention-LSTM prediction workflow.}
\label{fig:4}
\end{figure*}

Figure~\ref{fig:3} summarizes the regression process. The targets are
vortex-method power, thrust, and torque at the sampled time nodes.
For each configuration, XGBoost predicts the response over 0 to 600 s.
CNN-Attention-LSTM then uses that predicted history to estimate the
response from 600 to 1200 s. Figure~\ref{fig:4} shows this serial
connection. Time in the two-stage prediction is measured after removal
of the startup interval. For each evaluation condition, the second
stage is trained on the first 600 s and predicts the following 600 s by
free-running recursion. At each step, its predictions of all time-varying
input columns are fed back as inputs for the next step. No simulated
value from the second interval is used. This chronological split is the 50\% training
proportion in
Section~\ref{subsec:multi-parameter-power-prediction-for-twin-rotor-wind-turbines}.
The comparison uses either the MRUT history as the reference input or
the XGBoost prediction as the hybrid input for the first interval. Each model is min--max
scaled with the range of its own training interval. The six evaluation
conditions were not used to select any hyperparameter. All histories
share one wind realization, so no split separates training and test data
by wind realization (Section~\ref{subsec:scope}).

The XGBoost input features are standardized as

\begin{equation}
x_{\mathrm{norm}}=\frac{x-\mu}{\sigma}.
\label{eq:standardization}
\end{equation}

where $\mu$ and $\sigma$ are the mean and standard deviation of each
feature over the training configurations. The temporal network uses
min--max scaling instead (Section~\ref{subsec:lstm-model-verification}). With $y_i$ denoting the response target and $\theta$ the model
parameters, the training objective and prediction can be written as

\begin{equation}
\operatorname{Obj}(\theta)=\sum_{i=1}^{n}L(y_i,\widehat{y}_i)
+\sum_{k=1}^{K}\Omega(f_k).
\label{eq:training-objective}
\end{equation}

\begin{equation}
\widehat{y}_{\mathrm{XGBoost}}=f_{\mathrm{XGBoost}}(x_{\mathrm{norm}}).
\label{eq:xgb-prediction}
\end{equation}

For power prediction, the XGBoost response history is denoted by
$P_{\mathrm{XGBoost}}(t)$. The temporal processing stage is summarized
as

\begin{equation}
\left\{
\begin{aligned}
E_t^{\mathrm{input}}&=\operatorname{CNN}(P_{\mathrm{XGBoost}}(t)),\\
A_t^{\mathrm{input}}&=\operatorname{Attention}(E_t^{\mathrm{input}}),\\
h_t^{\mathrm{LSTM}}&=\operatorname{LSTM}(A_t^{\mathrm{input}},h_{t-1}^{\mathrm{LSTM}}),\\
\widehat{y}_{\mathrm{LSTM}}(t)&=\operatorname{Dense}(h_t^{\mathrm{LSTM}}).
\end{aligned}
\right.
\label{eq:lstm-stage}
\end{equation}

The second-stage output covers the subsequent 600 s interval. The
complete response is assembled as

\begin{equation}
\widehat{y}_{\mathrm{final}}=
\widehat{y}_{\mathrm{XGBoost}}\oplus\widehat{y}_{\mathrm{LSTM}}.
\label{eq:fusion}
\end{equation}

where $\oplus$ denotes temporal concatenation of the two consecutive
prediction intervals.

\subsection{Configuration optimization using a genetic algorithm}\label{subsec:twin-rotor-configuration-parameter-optimization-model-based-on-ga}

The GA searches combinations of tip-speed ratio and spacing ratio
without requiring response gradients. For compactness, let
$s=L/D$. For each candidate $(\lambda,s)$, the first-stage surrogate
returns the rotor histories $\widehat{P}_{m,k}(\lambda,s)$ and
$\widehat{T}_{m,k}(\lambda,s)$. These contain $N=4236$ time nodes in a
600 s window, with $m=1,2$ denoting WT1 and WT2. The configuration
parameters remain fixed within each response history. The
optimization objectives are the time-mean system power coefficient
$\overline{C}_P$ and the time-mean system thrust coefficient
$\overline{C}_T$ defined in
Section~\ref{subsec:twin-rotor-model-and-the-vortex-method}:

\begin{equation}
\left\{
\begin{aligned}
\overline{C}_P(\lambda,s)&=\frac{1}{N}\sum_{k=1}^{N}\frac{\widehat{P}_{1,k}(\lambda,s)+\widehat{P}_{2,k}(\lambda,s)}{\tfrac{\pi}{8}\rho U_\infty^3D_1^2},\\
\overline{C}_T(\lambda,s)&=\frac{1}{N}\sum_{k=1}^{N}\frac{\widehat{T}_{1,k}(\lambda,s)+\widehat{T}_{2,k}(\lambda,s)}{\tfrac{\pi}{8}\rho U_\infty^2D_1^2},\\
\text{objectives:}\quad&\max_{(\lambda,s)\in\Omega}\overline{C}_P(\lambda,s),\quad
\min_{(\lambda,s)\in\Omega}\overline{C}_T(\lambda,s).
\end{aligned}
\right.
\label{eq:objectives}
\end{equation}

At a given wind speed, $\overline{C}_P$ and $\overline{C}_T$ are
proportional to the mean power and the mean thrust of the pair. The mean thrust
measures the sustained aerodynamic load on the shared support. Peak,
fatigue, and platform loads are not part of the objectives
(Section~\ref{subsec:scope}). A configuration is nondominated if no other
feasible configuration provides at least as large a power coefficient
and no larger thrust coefficient, with a strict improvement in at least one
objective:

\begin{equation}
\begin{aligned}
\mathcal{S}_{\mathrm{Pareto}}=\{(\lambda,s)\in\Omega\mid&\nexists(\lambda',s')\in\Omega\ \text{s.t.}\\
&\overline{C}_P(\lambda',s')\geq\overline{C}_P(\lambda,s),\\
&\overline{C}_T(\lambda',s')\leq \overline{C}_T(\lambda,s),\\
&\text{with at least one strict inequality}\}.
\end{aligned}
\label{eq:pareto-set}
\end{equation}

The feasible design domain is $\Omega=[4,11]\times[0.01,0.20]$ in
$(\lambda,s)$. Equation~\eqref{eq:pareto-set} defines the target
Pareto set, which the GA approximates within the prescribed bounds.

\begin{figure*}[!t]
\centering
\includegraphics[width=0.95\linewidth,height=0.62\textheight,keepaspectratio]{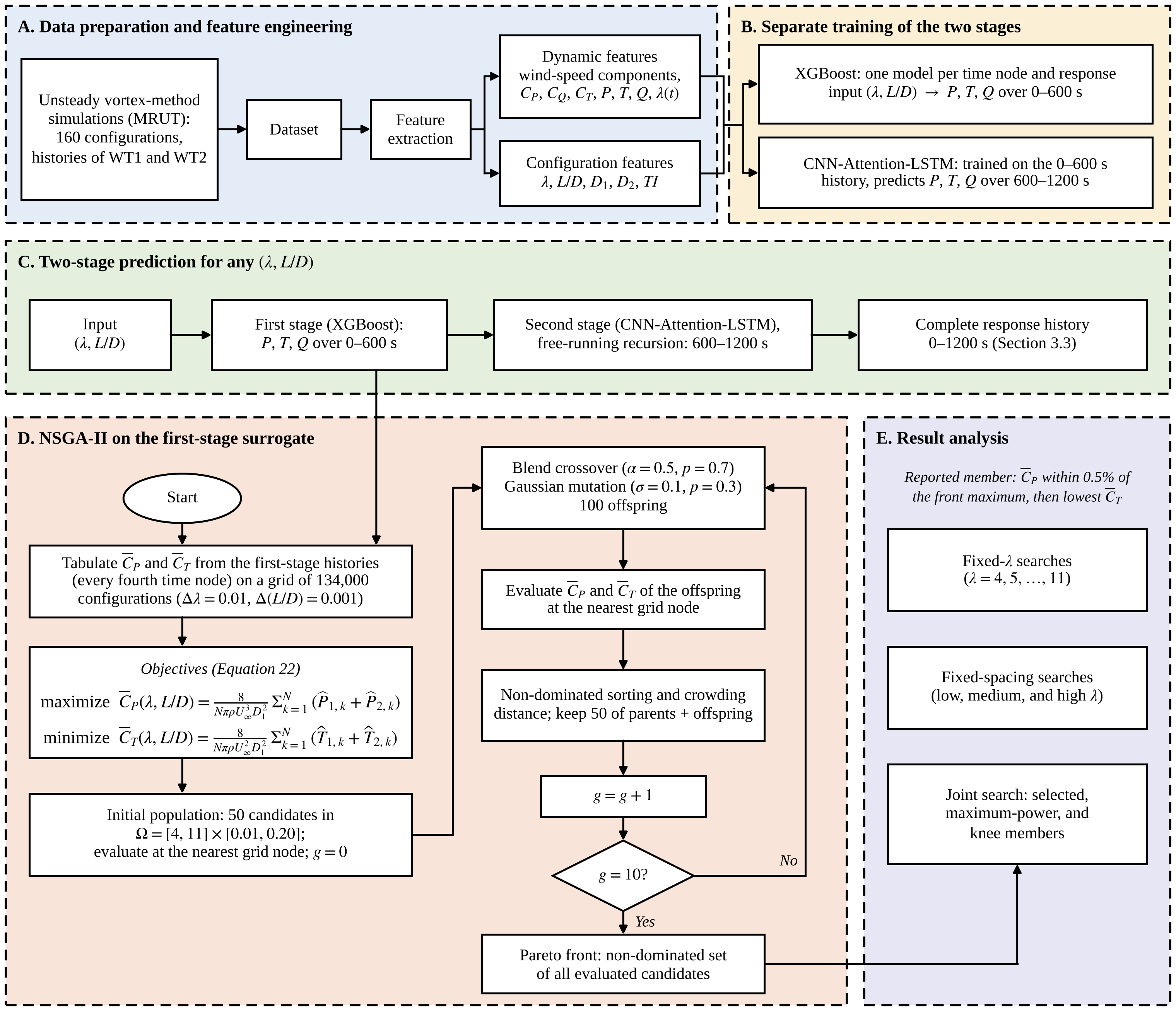}
\caption{Workflow of the data preparation, the two-stage prediction, and the configuration search. The GA evaluates the objectives of Equation~\eqref{eq:objectives} from the first-stage histories.}
\label{fig:5}
\end{figure*}

Figure~\ref{fig:5} summarizes five stages. Phase A generates MRUT
response histories and organizes the dynamic and configuration
features. Phase B trains the time-node XGBoost models and the temporal
network separately. Phase C connects them in series to predict a
1200 s response for a specified $(\lambda,L/D)$ pair.

In Phase D, the GA initializes a population of real-valued
$(\lambda,L/D)$ pairs within the prescribed bounds. The first-stage
surrogate predicts the power and thrust histories for each candidate,
from which the objectives in Equation~\eqref{eq:objectives} are
evaluated. Selection, crossover, and mutation generate subsequent
populations until the stopping criterion is reached. The GA follows
the NSGA-II scheme \cite{Deb2002} as implemented in DEAP
\cite{Fortin2012}. The population contains 50 individuals, with 50
parents and 100 offspring per generation under elitist plus-selection.
Blend crossover uses coefficient 0.5 and probability 0.7. Gaussian
mutation uses standard deviation 0.1, per-gene probability 0.2, and
probability 0.3. Selection uses non-dominated sorting and crowding
distance, and candidates are clipped to the bounds. The search stops after 10 generations, and the
non-dominated set of all evaluated candidates forms the reported front.
For the searches in
Section~\ref{subsec:parameter-optimization-study-for-twin-rotor-configuration},
the first-stage models were retrained on all 160 configurations with the
cross-validation settings of Section~\ref{subsec:cv}. Each rotor,
response, and time node has a model with 100 trees, maximum depth 6,
and learning rate 0.3. $\overline{C}_P$ and $\overline{C}_T$ were tabulated on a grid of 134,000
configurations ($\Delta\lambda=0.01$, $\Delta(L/D)=0.001$) from every
fourth time node, and each candidate is evaluated at the nearest grid
node. In the MRUT data, using every fourth node changes $\overline{C}_P$ by
at most 0.17\% and $\overline{C}_T$ by at most 0.04\%.

Phase E compares selected spacings at fixed tip-speed ratios, selected
tip-speed ratios at fixed spacings, and the front of the joint
search. Members whose $\overline{C}_P$ lies within 0.5\% of the front
maximum are treated as equal in power. Among these, the member with
the lowest $\overline{C}_T$ is reported. The tolerance equals the 90th
percentile of the cross-validated error of $\overline{C}_P$
(Section~\ref{subsec:cv}), so smaller power differences are not resolved
by the surrogate. For the joint search, the maximum-power member and the
knee of the front are also reported
(Section~\ref{subsec:parameter-optimization-study-for-twin-rotor-configuration}).
These comparisons show how the preferred configuration changes with the
permitted operating range.

\section{Results and Discussion}\label{sec:verification-and-discussion}

\subsection{LSTM model verification}\label{subsec:lstm-model-verification}

\begin{figure*}[!t]
\centering
\includegraphics[width=0.94\linewidth,height=0.57\textheight,keepaspectratio]{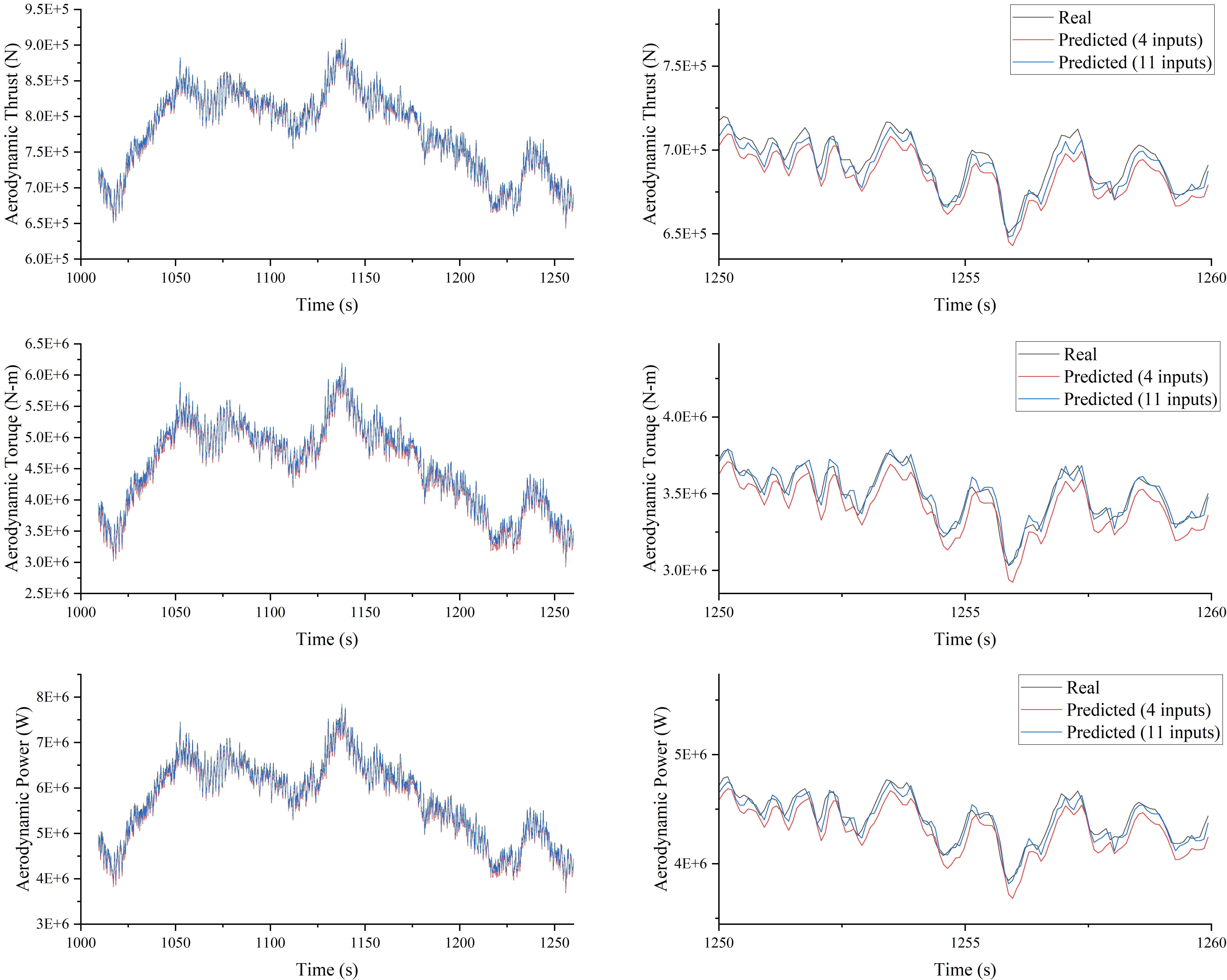}
\caption{LSTM predictions using four-column and eleven-column datasets. Both include a time column and contain three and ten non-time variables, respectively.}
\label{fig:6}
\end{figure*}

\begin{figure*}[!t]
\centering
\includegraphics[width=0.94\linewidth,height=0.57\textheight,keepaspectratio]{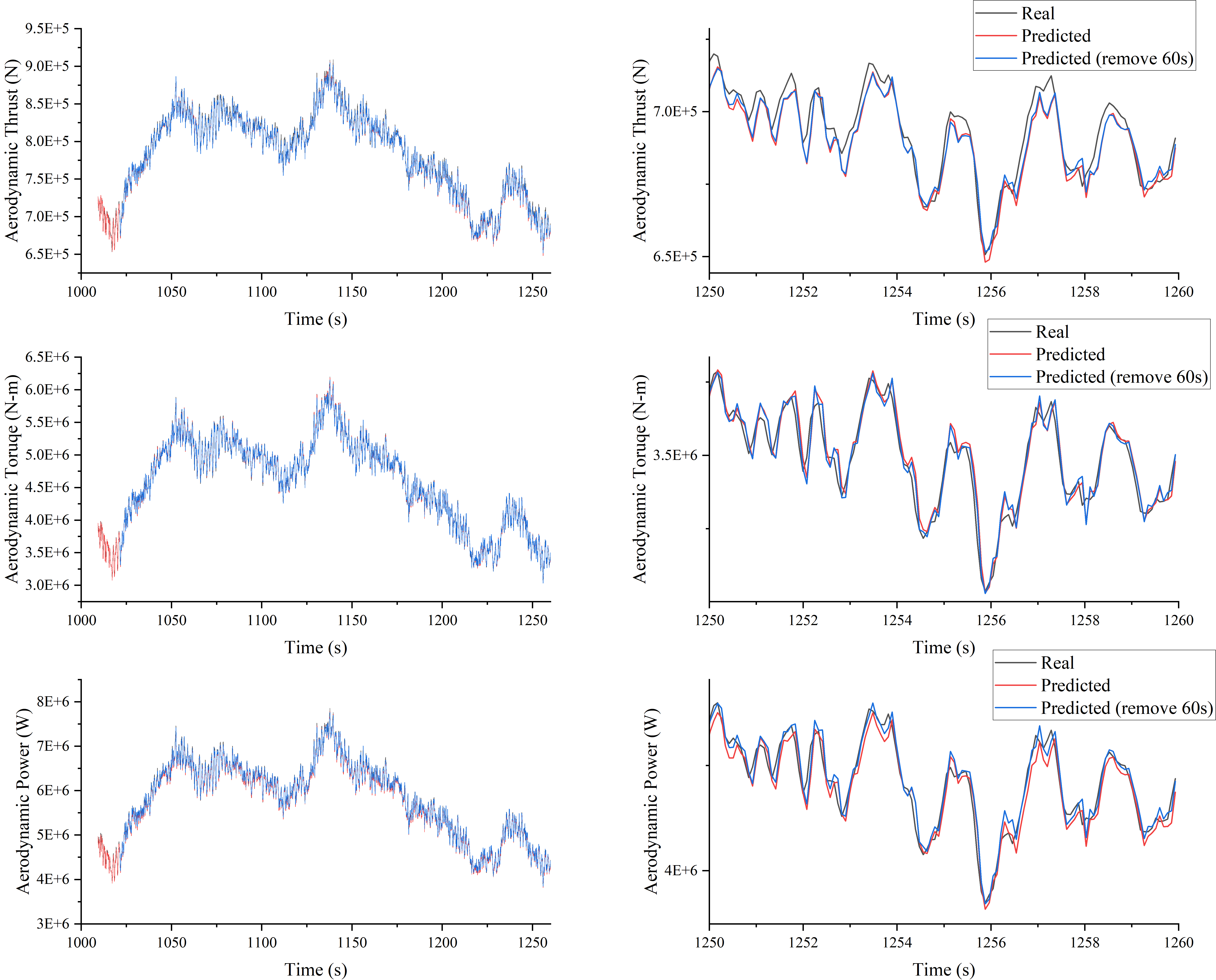}
\caption{Comparison of predictions before and after removing the initial transient period.}
\label{fig:7}
\end{figure*}

\begin{figure*}[!t]
\centering
\includegraphics[width=0.90\linewidth,height=0.57\textheight,keepaspectratio]{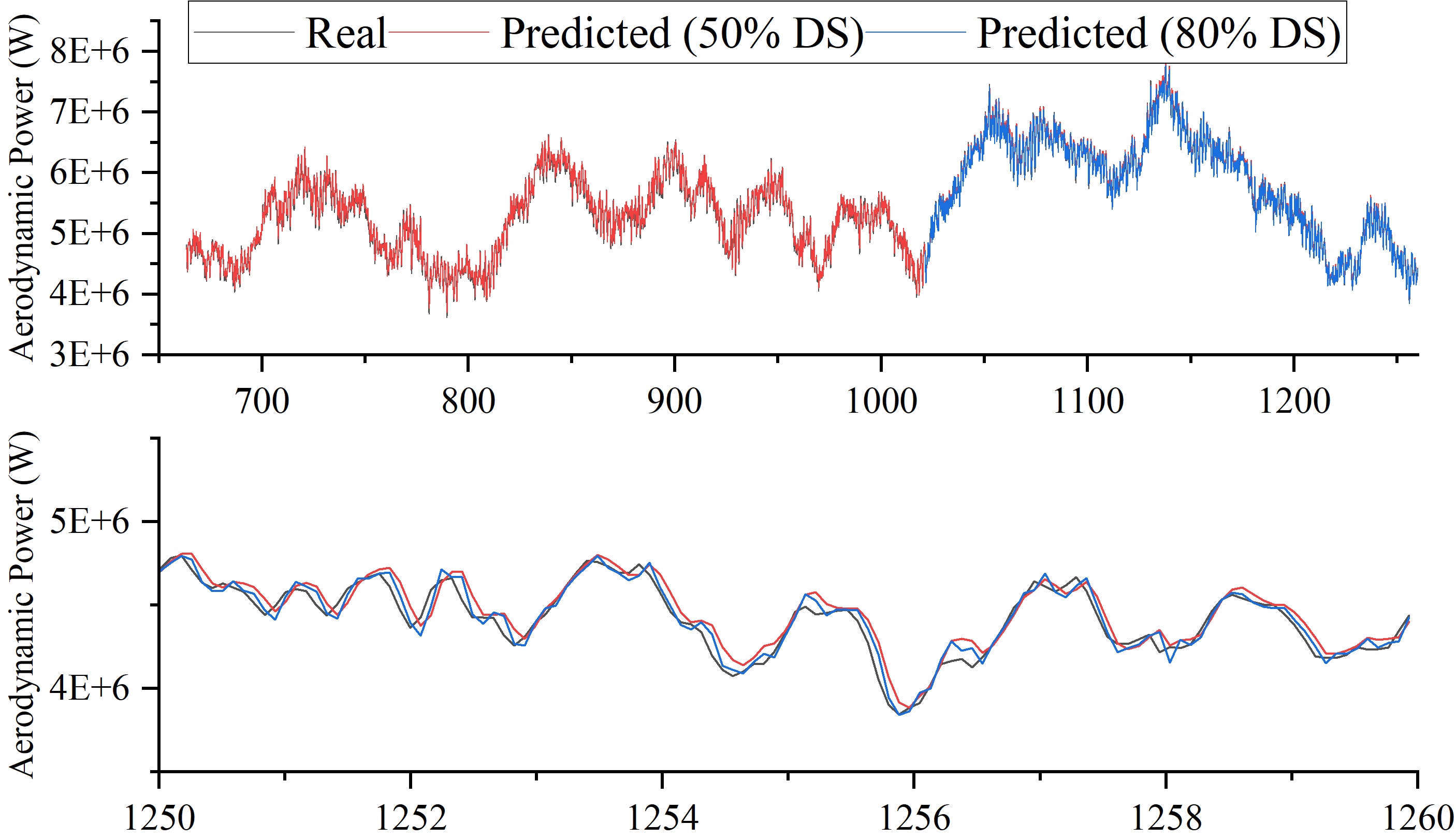}
\caption{Aerodynamic-power prediction using different training-set sizes.}
\label{fig:8}
\end{figure*}

The training dataset for the LSTM model was generated from 1260 s
time-domain vortex-method simulations. TurbSim was used to generate
unsteady wind speeds, with a turbulent wind field characterized over a
600 s period. Two complete turbulent-wind cycles were therefore
considered, together with an additional 60 s startup transient. The time
step was specified through a 6-degree azimuthal increment. At
12.1 rpm, this corresponds to $\Delta t=6/(6\times12.1)=0.0826$ s.
The verification case used a 11.4 m/s mean wind speed and 10\% turbulence
intensity. The time axes in Figs.~\ref{fig:6} to \ref{fig:8} retain the
original simulation clock. The two-stage prediction intervals are
measured relative to the end of startup.

The verification network uses a look-back window of 80 steps
(LOOKBACK) and a batch size of 64 (BATCH\_SIZE). Its two LSTM layers
contain 128 and 64 units. The forecast horizon is one step
(FORECAST\_SIZE=1), with 10 epochs (EPOCHS) and a training ratio of 0.8
(TRAIN\_RATIO). The look-back window of
80 steps spans 6.6 s. Each column is min--max scaled with the range of
the training portion, the training and test portions are separated
chronologically, and the network predicts one step ahead. The settings
were selected in two steps. An $L_{18}$ orthogonal
design varied the look-back window (40, 60, 80 steps), batch size (32,
64, 128), and LSTM units (64, 128, 256) on the verification case.
It also varied forecast horizon (1, 2, 3 steps) and epoch count
(5, 10, 15).
The look-back window, number of units, and forecast horizon had the
largest effect. Increasing the number of units to 256 raised training
time to 2125 s without a corresponding gain in accuracy. The final
setting was chosen in a nine-case study with $\lambda=6$, 8, 10 and
$L/D=0.05$, 0.10, 0.15. It raised $R^2$ in most cases relative to the
best setting of the orthogonal design.

The input comparison uses four-column and eleven-column datasets,
each with time in the first column. The four-column dataset contains
thrust, torque, and power as its three non-time variables. The
eleven-column dataset contains ten non-time variables, comprising three
wind-speed components, power, thrust, torque, the diameters of WT1 and
WT2, turbulence intensity, and spacing ratio. Both configurations
predict thrust, torque, and power.

The eleven-column dataset gives closer agreement with the simulation
histories in Figure~\ref{fig:6}. Table~\ref{tab:metrics} shows that the
MSE decreases from $5.19\times10^{-4}$ to $2.14\times10^{-4}$ for
thrust, from $3.69\times10^{-4}$ to $9.30\times10^{-5}$ for torque,
and from $5.50\times10^{-4}$ to $1.52\times10^{-4}$ for power. The
corresponding reductions are 58.8\%, 74.8\%, and 72.4\%. The expanded
input set improves prediction of all three responses in this case.
Within one history, the four configuration variables of the
eleven-column dataset are constant, so the inflow components are the
only additional time-varying input. A single comparison does not
separate their effect from training variability. The controlled
ablation in Section~\ref{subsec:benchmark} separates the input groups.

\begin{table*}[!t]
\caption{LSTM evaluation metrics for the input and training-data comparisons. Column counts include time.}
\label{tab:metrics}
\centering
\small\setlength{\tabcolsep}{4pt}\renewcommand{\arraystretch}{1.10}
\begin{tabular}{lccccc}
\toprule
\textbf{Metric} & \textbf{Response} & \textbf{4-column dataset} & \textbf{11-column dataset} & \textbf{Startup excluded} & \textbf{50\% training set} \\
\midrule
MSE & AeroThrust & 5.19E-04 & 2.14E-04 & 3.15E-04 & 7.65E-04 \\
MSE & AeroTorque & 3.69E-04 & 9.30E-05 & 1.70E-04 & 5.07E-04 \\
MSE & AeroPower & 5.50E-04 & 1.52E-04 & 1.80E-04 & 3.68E-04 \\
RMSE & AeroThrust & 2.28E-02 & 1.46E-02 & 1.78E-02 & 2.77E-02 \\
RMSE & AeroTorque & 1.92E-02 & 9.64E-03 & 1.31E-02 & 2.25E-02 \\
RMSE & AeroPower & 2.34E-02 & 1.23E-02 & 1.34E-02 & 1.92E-02 \\
MAE & AeroThrust & 1.93E-02 & 1.17E-02 & 1.41E-02 & 2.24E-02 \\
MAE & AeroTorque & 1.65E-02 & 7.54E-03 & 1.02E-02 & 1.84E-02 \\
MAE & AeroPower & 2.08E-02 & 1.00E-02 & 1.05E-02 & 1.53E-02 \\
$R^2$ & AeroThrust & 9.81E-01 & 9.92E-01 & 9.92E-01 & 9.78E-01 \\
$R^2$ & AeroTorque & 9.81E-01 & 9.95E-01 & 9.95E-01 & 9.84E-01 \\
$R^2$ & AeroPower & 9.71E-01 & 9.92E-01 & 9.95E-01 & 9.89E-01 \\
\bottomrule
\end{tabular}
\end{table*}

Subsequent predictions use the eleven-column dataset with the initial
60 s startup interval excluded. Figure~\ref{fig:7} shows the initial
oscillations and the predictions after their removal. The analysis
therefore focuses on post-startup operation. Relative to the full eleven-column
dataset, thrust, torque, and power MSE increase to
$3.15\times10^{-4}$, $1.70\times10^{-4}$, and
$1.80\times10^{-4}$, respectively. The corresponding $R^2$ values
are unchanged at the reported precision for thrust and torque and
increase from 0.992 to 0.995 for power. Startup removal thus changes
the evaluation window and raises MSE while retaining similar $R^2$ values.
Because $R^2=1-\mathrm{MSE}/\sigma^2$, where $\sigma^2$ is the
variance of the scaled target in the test window, a change in $\sigma^2$ allows MSE to increase while $R^2$ remains
similar. The values in
Table~\ref{tab:metrics} imply that $\sigma^2=\mathrm{MSE}/(1-R^2)$
increased by factors of about 1.5 (thrust), 1.8 (torque), and 1.9
(power) after trimming. The startup transient contains the largest
excursions of each history. Removing it narrows the min--max scaling
range and enlarges the scaled fluctuations, increasing both scaled MSE
and $\sigma^2$.

After startup removal, the training fraction was reduced from 80\%
to 50\%, with the remaining samples used for evaluation.
Figure~\ref{fig:8} compares the power predictions for these splits.

Reducing the training fraction increases the MSE for all three
responses, while the reported $R^2$ values remain at or above 0.978.
The model retains the main temporal pattern with half of the samples
used for training. The nine grid histories in
Section~\ref{subsec:benchmark} also test an intermediate training
fraction of 65\%. At fractions of 50\%, 65\%, and 80\%, the mean
one-step MSE values for WT2 are $3.5$, $3.2$, and
$3.0\times10^{-3}$, respectively. The corresponding WT1 values are
$1.8$, $1.8$, and $1.5\times10^{-3}$. Mean $R^2$ changes by at most
0.011 because the test window and target variance change with the
training fraction. The subsequent analysis uses the eleven-column
dataset after startup removal to maintain a consistent post-startup
response window.

\subsection{Comparison with simpler temporal models and input ablation}\label{subsec:benchmark}

The CNN-Attention-LSTM was compared with a vanilla LSTM, a
CNN-LSTM without attention, and persistence under the same split and
metrics. The vanilla LSTM uses the same two layers of 128 and 64 units.
Persistence repeats the last known value. The three networks share one
code implementation. Where present, the convolutional layer uses 64
filters and kernel size 3. Training uses the look-back window, batch
size, and epoch count from
Section~\ref{subsec:lstm-model-verification}, with the same random seed.
The tests use the 600 s MRUT histories of the nine grid configurations of
the nine-case study ($\lambda=6$, 8, 10 and $L/D=0.05$, 0.10, 0.15). At the
grid time step of 0.1417 s, the 80-step look-back window spans 11.3 s.
Each network receives the responses, inflow components, aerodynamic
coefficients, and instantaneous tip-speed ratio of one rotor. It is
trained on the first 80\% of the history and evaluated on the remaining
20\%. Each prediction is one or ten steps ahead, using the simulated
inputs of the preceding window. Inputs and targets are min--max scaled
with the training portion. The
benchmark networks process the inputs of one rotor without the
configuration and vortex-method branch of Equation~\eqref{eq:wt2-branch},
so they compare temporal architectures rather than the complete WT2
model. Because the nine configurations were also used to select the
settings in Section~\ref{subsec:lstm-model-verification}, the comparison
is made under these settings and not on unseen conditions.

\begin{table*}[!t]
\caption{One-step-ahead and ten-step-ahead $R^2$ of the temporal models, averaged over nine grid configurations ($\lambda=6$, 8, 10; $L/D=0.05$, 0.10, 0.15). Training on the first 80\% of each 600 s history; time step 0.1417 s. Persistence repeats the last known value.}
\label{tab:bench}
\centering
\small\setlength{\tabcolsep}{4pt}\renewcommand{\arraystretch}{1.10}
\begin{tabular}{lcccccc}
\toprule
 & \multicolumn{3}{c}{\textbf{WT1}} & \multicolumn{3}{c}{\textbf{WT2}} \\
\cmidrule(lr){2-4}\cmidrule(lr){5-7}
\textbf{Model} & \textbf{Thrust} & \textbf{Torque} & \textbf{Power} & \textbf{Thrust} & \textbf{Torque} & \textbf{Power} \\
\midrule
\multicolumn{7}{l}{\textit{One step ahead (0.14 s)}} \\
Persistence & 0.966 & 0.967 & 0.967 & 0.892 & 0.880 & 0.880 \\
LSTM & 0.941 & 0.956 & 0.954 & 0.921 & 0.914 & 0.915 \\
CNN-LSTM & 0.950 & 0.962 & 0.961 & 0.932 & 0.927 & 0.927 \\
CNN-Attention-LSTM & 0.960 & 0.970 & 0.970 & 0.930 & 0.925 & 0.926 \\
\midrule
\multicolumn{7}{l}{\textit{Ten steps ahead (1.42 s)}} \\
Persistence & 0.760 & 0.722 & 0.722 & 0.506 & 0.446 & 0.446 \\
LSTM & 0.885 & 0.875 & 0.877 & 0.907 & 0.902 & 0.902 \\
CNN-LSTM & 0.924 & 0.931 & 0.932 & 0.911 & 0.907 & 0.907 \\
CNN-Attention-LSTM & 0.915 & 0.914 & 0.913 & 0.914 & 0.909 & 0.908 \\
\bottomrule
\end{tabular}
\end{table*}

One step ahead (0.14 s), persistence gives mean $R^2$ values of
0.966--0.967 for WT1 and 0.880--0.892 for WT2
(Table~\ref{tab:bench}). For WT1, the vanilla LSTM falls below
persistence for every configuration and response. The convolutional
layer raises mean $R^2$ by 0.007, and attention adds 0.009. This brings
CNN-Attention-LSTM level with persistence, exceeding it in 16 of 27
configuration--response combinations. For WT2, all three networks
exceed persistence by 0.03--0.04 on average. Convolution adds 0.012 to
the vanilla LSTM, and attention changes mean $R^2$ by $-0.001$.
One-step metrics therefore mainly reflect the histories' strong
short-term autocorrelation. Ten steps ahead (1.42 s), persistence
degrades to mean $R^2$ values of 0.73 for WT1 and 0.47 for WT2.
The three networks retain 0.88--0.93 and exceed persistence for every
configuration and response. At this horizon, convolution raises mean
$R^2$ over the vanilla LSTM by 0.050 for WT1 and 0.004 for WT2.
Adding attention lowers it by 0.015 for WT1 and raises it by 0.002 for
WT2. Convolution accounts for most of the gain at both horizons.
Attention benefits only one-step prediction for WT1. The median
training time per history was 23 s for the LSTM, 26 s for the CNN-LSTM,
and 28 s for the CNN-Attention-LSTM.

\begin{table*}[!t]
\caption{Input ablation of CNN-Attention-LSTM using one-step-ahead $R^2$ averaged over the nine configurations of Table~\ref{tab:bench}. Each rotor uses its own histories. The last row adds the WT1 histories to the WT2 inputs.}
\label{tab:ablation}
\centering
\small\setlength{\tabcolsep}{4pt}\renewcommand{\arraystretch}{1.10}
\begin{tabular}{lcccccc}
\toprule
 & \multicolumn{3}{c}{\textbf{WT1}} & \multicolumn{3}{c}{\textbf{WT2}} \\
\cmidrule(lr){2-4}\cmidrule(lr){5-7}
\textbf{Inputs} & \textbf{Thrust} & \textbf{Torque} & \textbf{Power} & \textbf{Thrust} & \textbf{Torque} & \textbf{Power} \\
\midrule
Responses (R) & 0.961 & 0.965 & 0.965 & 0.923 & 0.918 & 0.918 \\
R + inflow components (W) & 0.958 & 0.970 & 0.969 & 0.916 & 0.913 & 0.913 \\
R + W + coefficients (C) & 0.957 & 0.966 & 0.966 & 0.927 & 0.923 & 0.923 \\
R + W + C + tip-speed ratio $\lambda(t)$ & 0.960 & 0.970 & 0.970 & 0.930 & 0.925 & 0.926 \\
All WT2 inputs + all WT1 inputs & -- & -- & -- & 0.923 & 0.918 & 0.918 \\
\bottomrule
\end{tabular}
\end{table*}

Table~\ref{tab:ablation} separates the input groups of the
CNN-Attention-LSTM one step ahead. For WT1, the four input sets give
mean $R^2$ values between 0.963 and 0.966. Adding inflow components,
coefficients, or tip-speed ratio to the responses changes the result by
at most 0.003. For WT2, the full input set gives the highest mean
$R^2$ of 0.927, compared with 0.920 using responses alone. The gain
mainly comes from aerodynamic coefficients and tip-speed ratio. Inflow
components alone lower it slightly to 0.914. Adding the WT1
histories to the WT2 inputs lowers the mean $R^2$ to 0.920, consistent
with the nearly identical inflow records of the two rotors
(Section~\ref{subsec:multi-parameter-power-prediction-for-twin-rotor-wind-turbines}).
One step ahead, the response histories therefore carry most of the
predictive information. The gain of the eleven-column dataset in
Table~\ref{tab:metrics}, obtained on the verification case, does not
appear as a consistent effect on the grid histories.

\subsection{Multi-parameter power prediction for twin-rotor wind turbines}\label{subsec:multi-parameter-power-prediction-for-twin-rotor-wind-turbines}

The simulation grid contains 20 spacing ratios, from 0.01 to 0.20 in
increments of 0.01, and eight tip-speed ratios, from 4 to 11 in
increments of one. Here, $L$ is the distance between rotor centers and
$D$ is the WT1 rotor diameter (284 m). The grid therefore contains 160
tandem configurations and 320 rotor-specific response sets.
Table~\ref{tab:rotor-speeds} lists the rotor
speeds, and Table~\ref{tab:hyperparameters} summarizes the model
settings.

These grid points define the training domain, $0.01\leq L/D\leq0.20$
and $4\leq\lambda\leq11$. Tables~\ref{tab:wt1-metrics} and
\ref{tab:wt2-metrics} report six evaluation conditions. Four lie within
the simulation-grid bounds, whereas $L/D=0.009$ and $\lambda=11.1$ lie
just outside the stated spacing and tip-speed-ratio ranges,
respectively, by 0.001 in $L/D$ and 0.1 in $\lambda$. None of the
six conditions is a grid point. Each was simulated separately with MRUT
for 1260 s. The latter two conditions test limited extrapolation. XGBoost predicts the first
600 s, with $R^2$ ranging from 0.934740 to 0.997783 for WT1 and from
0.908769 to 0.999216 for WT2. Four of the six cases exceed 0.99 for
each rotor, but the lower-scoring cases require closer examination
because their errors enter the temporal prediction stage.

\begin{table*}[!t]
\caption{Rotor rotational speeds of WT1 and WT2 under different tip-speed ratios at the simulated wind speed of 11.4 m/s, and the range of wind speeds below the WT1 rated wind speed of 11.0 m/s in which both rotors can run at the given tip-speed ratio (Figure~\ref{fig:envelope}). The rotor-speed ranges of the reference turbines are 1.97--7.061 rpm (WT1) \cite{ref31,IEA22data} and 6.9--12.1 rpm (WT2) \cite{ref32}.}
\label{tab:rotor-speeds}
\centering
\small\setlength{\tabcolsep}{4pt}\renewcommand{\arraystretch}{1.10}
\begin{tabular}{lcccccccc}
\toprule
\textbf{Tip-speed ratio, $\lambda$} & \textbf{4} & \textbf{5} & \textbf{6} & \textbf{7} & \textbf{8} & \textbf{9} & \textbf{10} & \textbf{11} \\
\midrule
WT1 rotational speed (rpm) & 3.1 & 3.8 & 4.6 & 5.4 & 6.1 & 6.9 & 7.6 & 8.4 \\
WT2 rotational speed (rpm) & 6.9 & 8.6 & 10.4 & 12.1 & 13.8 & 15.6 & 17.3 & 19 \\
Wind speed in envelope (m/s) & none & 9.1--11.0 & 7.6--11.0 & 6.5--11.0 & 5.7--10.0 & 5.1--8.9 & 4.6--8.0 & 4.1--7.3 \\
\bottomrule
\end{tabular}
\end{table*}

\begin{table*}[!t]
\caption{Hyperparameters of the XGBoost and LSTM models.}
\label{tab:hyperparameters}
\centering
\small\setlength{\tabcolsep}{4pt}\renewcommand{\arraystretch}{1.10}
\begin{tabular}{lcc}
\toprule
\textbf{Algorithm} & \textbf{LSTM} & \textbf{XGBoost} \\
\midrule
Parameters & \shortstack[l]{LOOKBACK=80\\BATCH\_SIZE=64\\LSTM\_UNITS=128/64 (two layers)\\EPOCHS=10\\FORECAST\_SIZE=1} & \shortstack[l]{Randomized search, 20 settings, three-fold CV:\\trees 100--400; learning rate 0.01--0.2;\\maximum depth 3--9; row and column\\sampling 0.7--1.0; $\gamma$ 0--0.2; L1/L2 penalties} \\
Model & \shortstack[l]{Hyperparameter optimization\\Static combined with dynamic features} & \shortstack[l]{Feature engineering: $L/D$, $\lambda$, rotor index,\\$\lambda L/D$, $\lambda^2$\\Random search} \\
\bottomrule
\end{tabular}
\end{table*}

The LSTM reference uses the simulated 0 to 600 s history to predict
the subsequent 600 s. This reference uses a simulated input history
without first-stage surrogate error. Its $R^2$ values range from 0.714194 to 0.979265
for WT1 and from 0.798111 to 0.951059 for WT2. Performance therefore
depends on the operating condition even when the input history comes
directly from the vortex simulation.

\begin{table*}[!t]
\caption{Comparison of WT1 prediction metrics across different algorithms. All conditions use 11.4 m/s mean wind speed and 10\% turbulence intensity. XGBoost is trained on the 160 grid configurations. The second stage is trained on the first 600 s of each condition. $(L/D,\lambda)=(0.009,4.0)$ and $(0.20,11.1)$ lie outside the grid. The other four lie inside it. Metrics are computed on min--max-scaled power. Each model is scaled
with the range of its own training interval, so MSE, RMSE, and MAE are
comparable within the rows of one model, whereas $R^2$ is unaffected by
the scaling.}
\label{tab:wt1-metrics}
\centering
\small\setlength{\tabcolsep}{4pt}\renewcommand{\arraystretch}{1.10}
\begin{tabular}{lcccccccc}
\toprule
\textbf{Algorithm} & \textbf{\shortstack{Training\\proportion}} & \textbf{\shortstack{Prediction\\interval}} & \textbf{$L/D$} & \textbf{$\lambda$} & \textbf{MSE} & \textbf{RMSE} & \textbf{MAE} & \textbf{$R^2$} \\
\midrule
XGBoost & 100\% & 0-600 s & 0.009 & 4.0 & 0.000102 & 0.010077 & 0.005743 & 0.995118 \\
XGBoost & 100\% & 0-600 s & 0.075 & 7.0 & 0.000123 & 0.011101 & 0.008805 & 0.997783 \\
XGBoost & 100\% & 0-600 s & 0.08 & 6.5 & 0.0038 & 0.061642 & 0.061147 & 0.93474 \\
XGBoost & 100\% & 0-600 s & 0.09 & 7.2 & 0.001351 & 0.036753 & 0.036316 & 0.975465 \\
XGBoost & 100\% & 0-600 s & 0.115 & 8.0 & 0.000156 & 0.012493 & 0.009781 & 0.997123 \\
XGBoost & 100\% & 0-600 s & 0.20 & 11.1 & 0.000367 & 0.019157 & 0.018834 & 0.99268 \\
LSTM & 50\% & 600-1200 s & 0.009 & 4.0 & 0.005852 & 0.076498 & 0.058886 & 0.714194 \\
LSTM & 50\% & 600-1200 s & 0.075 & 7.0 & 0.001114 & 0.033375 & 0.026616 & 0.979265 \\
LSTM & 50\% & 600-1200 s & 0.08 & 6.5 & 0.001488 & 0.038577 & 0.030827 & 0.972348 \\
LSTM & 50\% & 600-1200 s & 0.09 & 7.2 & 0.001592 & 0.039905 & 0.03215 & 0.970306 \\
LSTM & 50\% & 600-1200 s & 0.115 & 8.0 & 0.001319 & 0.036316 & 0.028819 & 0.974863 \\
LSTM & 50\% & 600-1200 s & 0.20 & 11.1 & 0.002672 & 0.051687 & 0.040622 & 0.939402 \\
XGBoost+LSTM & 50\% & 600-1200 s & 0.009 & 4.0 & 0.003979 & 0.063078 & 0.047828 & 0.804882 \\
XGBoost+LSTM & 50\% & 600-1200 s & 0.075 & 7.0 & 0.001112 & 0.033343 & 0.026489 & 0.979307 \\
XGBoost+LSTM & 50\% & 600-1200 s & 0.08 & 6.5 & 0.003065 & 0.055366 & 0.048634 & 0.941108 \\
XGBoost+LSTM & 50\% & 600-1200 s & 0.09 & 7.2 & 0.001132 & 0.033645 & 0.027156 & 0.977884 \\
XGBoost+LSTM & 50\% & 600-1200 s & 0.115 & 8.0 & 0.001206 & 0.034734 & 0.027568 & 0.976457 \\
XGBoost+LSTM & 50\% & 600-1200 s & 0.20 & 11.1 & 0.001475 & 0.038404 & 0.030578 & 0.968453 \\
\bottomrule
\end{tabular}
\end{table*}

Replacing the simulated history with the XGBoost prediction gives the
hybrid results in Tables~\ref{tab:wt1-metrics} and
\ref{tab:wt2-metrics}. The hybrid $R^2$ ranges are 0.804882 to
0.979307 for WT1 and 0.790232 to 0.957561 for WT2. Some hybrid results
improve on the simulation-history reference, while others deteriorate.
The effect of the predicted input history therefore depends on the
operating condition.

\begin{table*}[!t]
\caption{Comparison of WT2 prediction metrics across different algorithms. Test conditions and data split as in Table~\ref{tab:wt1-metrics}.}
\label{tab:wt2-metrics}
\centering
\small\setlength{\tabcolsep}{4pt}\renewcommand{\arraystretch}{1.10}
\begin{tabular}{lcccccccc}
\toprule
\textbf{Algorithm} & \textbf{\shortstack{Training\\proportion}} & \textbf{\shortstack{Prediction\\interval}} & \textbf{$L/D$} & \textbf{$\lambda$} & \textbf{MSE} & \textbf{RMSE} & \textbf{MAE} & \textbf{$R^2$} \\
\midrule
XGBoost & 100\% & 0-600 s & 0.009 & 4.0 & 0.000169 & 0.013007 & 0.005797 & 0.992044 \\
XGBoost & 100\% & 0-600 s & 0.075 & 7.0 & 0.00043 & 0.020748 & 0.015908 & 0.991156 \\
XGBoost & 100\% & 0-600 s & 0.08 & 6.5 & 0.004691 & 0.068493 & 0.066627 & 0.908769 \\
XGBoost & 100\% & 0-600 s & 0.09 & 7.2 & 0.000399 & 0.019978 & 0.019403 & 0.992674 \\
XGBoost & 100\% & 0-600 s & 0.115 & 8.0 & 0.00082 & 0.028635 & 0.022283 & 0.984441 \\
XGBoost & 100\% & 0-600 s & 0.20 & 11.1 & 2.80E-05 & 0.005289 & 0.004573 & 0.999216 \\
LSTM & 50\% & 600-1200 s & 0.009 & 4.0 & 0.002092 & 0.045739 & 0.032013 & 0.894012 \\
LSTM & 50\% & 600-1200 s & 0.075 & 7.0 & 0.002302 & 0.047979 & 0.038197 & 0.945635 \\
LSTM & 50\% & 600-1200 s & 0.08 & 6.5 & 0.002583 & 0.050823 & 0.039667 & 0.94489 \\
LSTM & 50\% & 600-1200 s & 0.09 & 7.2 & 0.002324 & 0.048208 & 0.038631 & 0.951059 \\
LSTM & 50\% & 600-1200 s & 0.115 & 8.0 & 0.003155 & 0.05617 & 0.044534 & 0.929891 \\
LSTM & 50\% & 600-1200 s & 0.20 & 11.1 & 0.00622 & 0.078866 & 0.062798 & 0.798111 \\
XGBoost+LSTM & 50\% & 600-1200 s & 0.009 & 4.0 & 0.001446 & 0.038029 & 0.025296 & 0.933396 \\
XGBoost+LSTM & 50\% & 600-1200 s & 0.075 & 7.0 & 0.001848 & 0.042993 & 0.033896 & 0.956328 \\
XGBoost+LSTM & 50\% & 600-1200 s & 0.08 & 6.5 & 0.005063 & 0.071154 & 0.060806 & 0.892671 \\
XGBoost+LSTM & 50\% & 600-1200 s & 0.09 & 7.2 & 0.002014 & 0.044882 & 0.035645 & 0.957561 \\
XGBoost+LSTM & 50\% & 600-1200 s & 0.115 & 8.0 & 0.002583 & 0.050824 & 0.040182 & 0.944507 \\
XGBoost+LSTM & 50\% & 600-1200 s & 0.20 & 11.1 & 0.006469 & 0.080431 & 0.063925 & 0.790232 \\
\bottomrule
\end{tabular}
\end{table*}

At $L/D=0.009$ and $\lambda=4.0$, WT1 hybrid $R^2$ is 0.805,
compared with 0.714 for the simulation-history reference.
For WT2 at $L/D=0.20$ and $\lambda=11.1$, both second-stage models
perform less well, with $R^2$ values of 0.798 and 0.790 for the
reference and hybrid predictions. XGBoost nevertheless gives
$R^2=0.999$ for the first stage. Accurate reconstruction of the first
interval is therefore not sufficient to ensure accurate prediction
of the second, particularly outside the trained tip-speed-ratio range.

WT1 has higher $R^2$ than WT2 in five of the six cases for both the
simulation-history reference and the hybrid model. The downstream
rotor is generally more difficult to predict in these tests. The
largest losses in accuracy occur at $L/D=0.08$, $\lambda=6.5$ and,
for WT2, at $L/D=0.20$, $\lambda=11.1$.

The reference and hybrid second stages use the same network
settings and test interval, but are trained on different first-interval
histories, each scaled with its own range. Their MSE values are
therefore on different scales, and the comparison uses $R^2$, which this
linear scaling does not change. The hybrid $R^2$ falls below the
reference by more than 0.01 only at $(0.08, 6.5)$. This condition has
the lowest first-stage $R^2$ for both rotors, 0.935 and 0.909.
The hybrid reductions are 0.031 for WT1 and 0.052 for WT2. In the other cases, the change ranges
from $-0.008$ to $+0.091$, with the largest gain at $(0.009, 4.0)$ for
WT1. In these cases, first-stage errors propagate where first-stage
accuracy is lowest. Where first-stage $R^2$ exceeds 0.97, hybrid $R^2$
is at most 0.008 below the reference.

Three observations are consistent with the lower WT2 accuracy,
although none of them was isolated. WT2 power fluctuates more strongly
relative to its mean. Across the 160 configurations, its coefficient of
variation averages 0.28, compared with 0.11 for WT1. At small spacing,
deep periodic drops are reproduced only in part
(Section~\ref{subsec:cv}). The recorded axial inflow means are similar
for WT2 and WT1, at 11.53 and 11.54 m/s, respectively. These records do
not contain the mean wake deficit. That deficit reaches the network
through WT2 responses and vortex-method descriptors. Both branches also use the same network size,
although the WT2 mapping also contains the upstream influence. The
first-stage cross-validation in Section~\ref{subsec:cv} shows the same
ordering.

\subsection{Cross-validated accuracy over the design space}\label{subsec:cv}

The six evaluation conditions test the hybrid model at a few
points. To map the first-stage accuracy over the whole design space, the
160 grid configurations were split into ten folds. Per-node XGBoost
models were trained on nine folds and predicted the remaining fold,
using every fourth time node. Each model uses 100 trees, maximum depth
6, and learning rate 0.3. Piecewise-linear interpolation between the
training configurations served as the simplest alternative surrogate.
Outside their convex hull, the nearest configuration was used. Further
tests withheld complete boundary rows of the grid ($\lambda=4$,
$\lambda=11$, $L/D\leq0.02$, or $L/D\geq0.19$), so that these
configurations had to be extrapolated. Accuracy is measured by $R^2$ and
by the normalized root-mean-square error (NRMSE, the RMSE divided by the
mean of the history) of each predicted power history.

\begin{figure*}[!t]
\centering
\includegraphics[width=\linewidth]{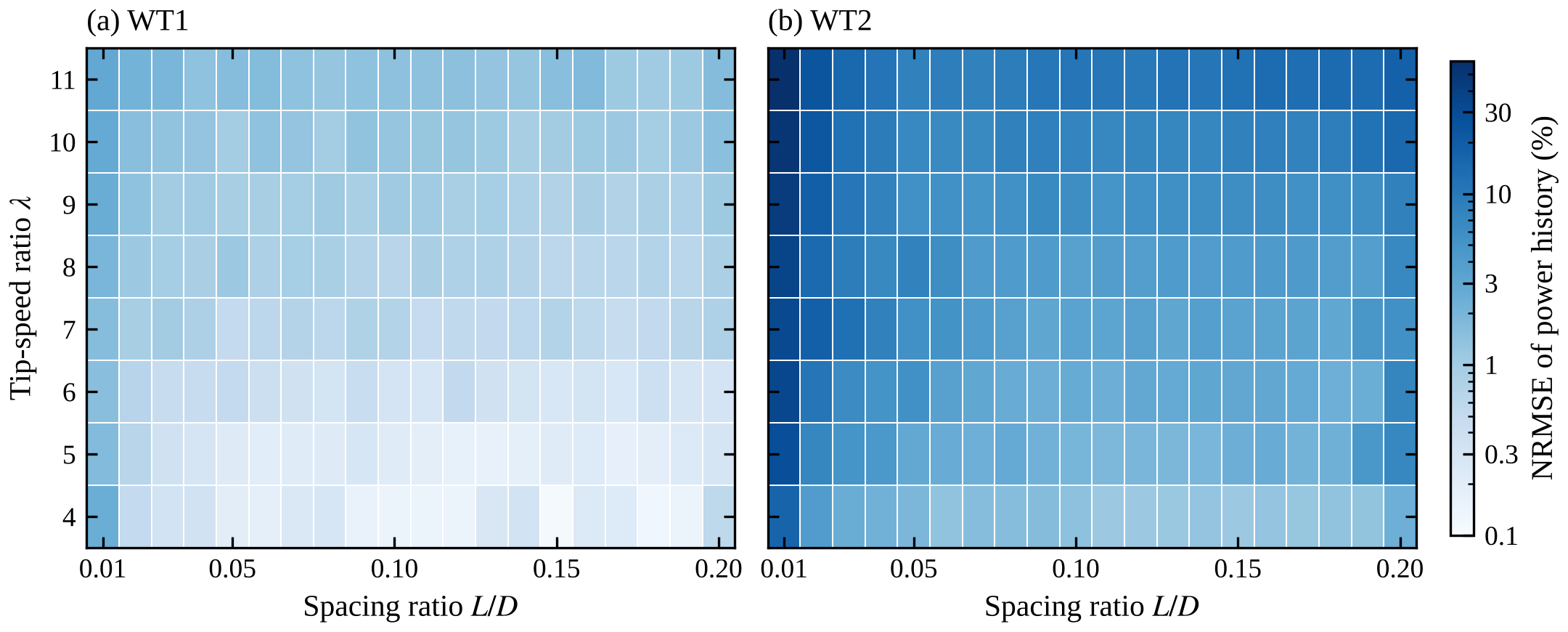}
\caption{Ten-fold cross-validated NRMSE of the first-stage power histories over the design space: (a) WT1; (b) WT2. Each cell is predicted by models that did not see that configuration.}
\label{fig:errormap}
\end{figure*}

\begin{figure*}[!t]
\centering
\includegraphics[width=\linewidth]{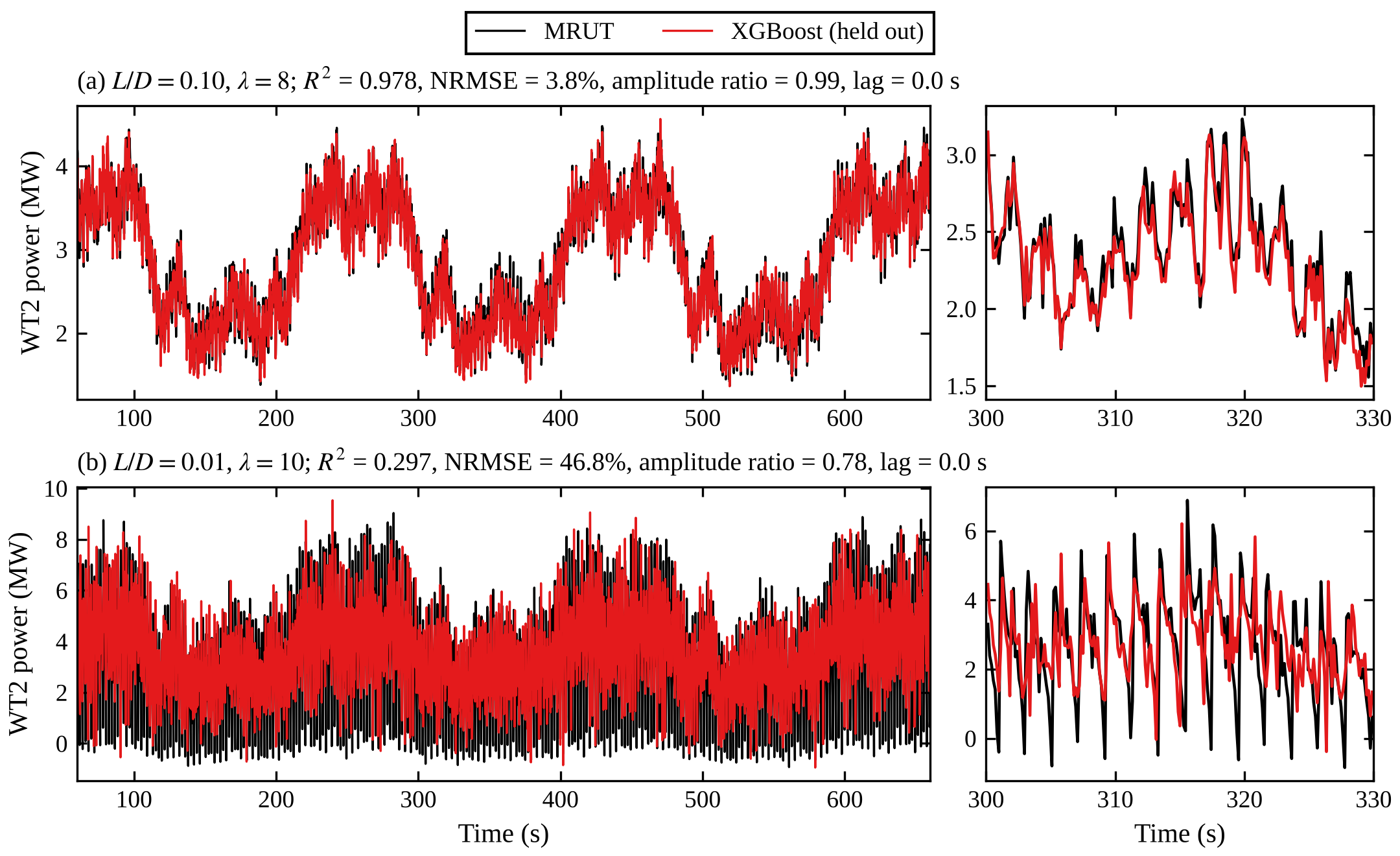}
\caption{MRUT and held-out first-stage predictions of the WT2 power history: (a) $L/D=0.10$, $\lambda=8$; (b) $L/D=0.01$, $\lambda=10$. The right panels enlarge 300--330 s. The prediction for each case comes from models trained on the other 159 configurations.}
\label{fig:trace}
\end{figure*}

For WT1, the median $R^2$ is 0.995 and the mean NRMSE is 0.8\%
(Figure~\ref{fig:errormap}a). For WT2, the median $R^2$ is 0.958 and the
mean NRMSE is 7.6\%. The error grows with $\lambda$ and peaks at
$L/D=0.01$ and 0.02 (Figure~\ref{fig:errormap}b). WT2 power changes
most steeply with spacing
(Section~\ref{subsec:parameter-optimization-study-for-twin-rotor-configuration}).
Figure~\ref{fig:trace} compares two WT2 histories at full time
resolution, each predicted by models trained on the other 159
configurations. At $L/D=0.10$ and $\lambda=8$, the prediction follows the
oscillations with an amplitude ratio of 0.99. This ratio divides the
prediction's standard deviation by the simulation's standard deviation.
The cross-correlation peaks at zero lag at the 0.14 s resolution. At
$L/D=0.01$ and $\lambda=10$, the history is dominated by deep periodic
power drops. The prediction keeps their timing (zero lag) but reproduces
only 78\% of the fluctuation amplitude ($R^2=0.30$).

\begin{table*}[!t]
\caption{First-stage accuracy for unseen configurations (power histories). Cross-validation (CV) uses ten folds over the 160 grid configurations. The interior excludes boundary rows. In boundary tests, the listed row is withheld from training. The CV value for the same configurations is given for comparison.}
\label{tab:cv}
\centering
\small\setlength{\tabcolsep}{4pt}\renewcommand{\arraystretch}{1.10}
\begin{tabular}{lcccc}
\toprule
 & \multicolumn{2}{c}{\textbf{WT1}} & \multicolumn{2}{c}{\textbf{WT2}} \\
\cmidrule(lr){2-3}\cmidrule(lr){4-5}
\textbf{Test} & \textbf{Median $R^2$} & \textbf{Mean NRMSE} & \textbf{Median $R^2$} & \textbf{Mean NRMSE} \\
\midrule
CV, XGBoost (default settings) & 0.995 & 0.82\% & 0.958 & 7.6\% \\
CV, XGBoost (tuned in each fold) & 0.996 & 0.86\% & 0.962 & 7.2\% \\
CV, linear interpolation & 0.997 & 1.1\% & 0.987 & 5.8\% \\
Interior, XGBoost (default settings) & 0.996 & 0.71\% & 0.969 & 5.5\% \\
Interior, linear interpolation & 0.998 & 0.62\% & 0.990 & 3.2\% \\
\midrule
 & \multicolumn{4}{c}{\textbf{Mean NRMSE, withheld row (CV value)}} \\
\cmidrule(lr){2-5}
$\lambda=4$ withheld & \multicolumn{2}{c}{87\% (0.36\%)} & \multicolumn{2}{c}{39\% (2.4\%)} \\
$\lambda=11$ withheld & \multicolumn{2}{c}{7.0\% (1.5\%)} & \multicolumn{2}{c}{42\% (15\%)} \\
$L/D\leq0.02$ withheld & \multicolumn{2}{c}{1.7\% (1.7\%)} & \multicolumn{2}{c}{19\% (26\%)} \\
$L/D\geq0.19$ withheld & \multicolumn{2}{c}{0.86\% (0.76\%)} & \multicolumn{2}{c}{8.3\% (7.4\%)} \\
\bottomrule
\end{tabular}
\end{table*}

Table~\ref{tab:cv} summarizes the tests. Tuning the XGBoost
hyperparameters by randomized search within each training fold changes
the WT2 median $R^2$ only from 0.958 to 0.962. Linear interpolation gives
a higher median $R^2$ for both rotors and lower mean NRMSE for WT2.
In the grid interior, WT2 NRMSE is 3.2\% with interpolation and 5.5\%
with XGBoost. XGBoost has lower mean NRMSE for WT1, at 0.82\% against
1.1\%. On this
regular two-parameter grid, the configuration-to-history mapping is
smooth enough for interpolation, and the tree ensemble does not improve
on it. The regression model can accommodate further inputs, such as
wind speed and turbulence intensity, but its advantage over
interpolation in a higher-dimensional design space remains to be
tested. Withholding boundary rows
exposes the extrapolation limit. Without the $\lambda=11$ row, the mean
NRMSE at $\lambda=11$ rises from 1.5\% to 7.0\% for WT1 and from 15\% to
42\% for WT2. Without the $\lambda=4$ row, it reaches 87\% for WT1 and
39\% for WT2, because the tree ensemble returns the response of the
nearest sampled tip-speed ratio. Withholding a spacing boundary changes
the WT1 error by at most 0.1 percentage points. The out-of-range
conditions in Tables~\ref{tab:wt1-metrics} and \ref{tab:wt2-metrics} lie
within 0.1 in $\lambda$ and 0.001 in $L/D$ of the grid and are therefore
only slight extrapolations. The surrogate should not be used outside the
sampled range of $\lambda$.

The learning-curve test held out 32 configurations. Models were
trained on the remaining 128 or on three random subsets of 96, 64, and
32 configurations. Mean power NRMSE for WT1 changes from 1.3\% with
32 configurations to 0.81\% with 96 and 0.76\% with 128. The
corresponding WT2 values decrease from 12\% to 7.5\% and 6.8\%. The
WT1 error decreases only slightly between 96 and 128 configurations,
whereas the WT2 error continues to decrease. For WT2, additional samples
at $L/D\leq0.03$ and high $\lambda$, where its error is concentrated,
would be most useful.

\subsection{Parameter optimization study for twin-rotor configuration}\label{subsec:parameter-optimization-study-for-twin-rotor-configuration}

\begin{figure*}[!t]
\centering
\includegraphics[width=0.94\linewidth]{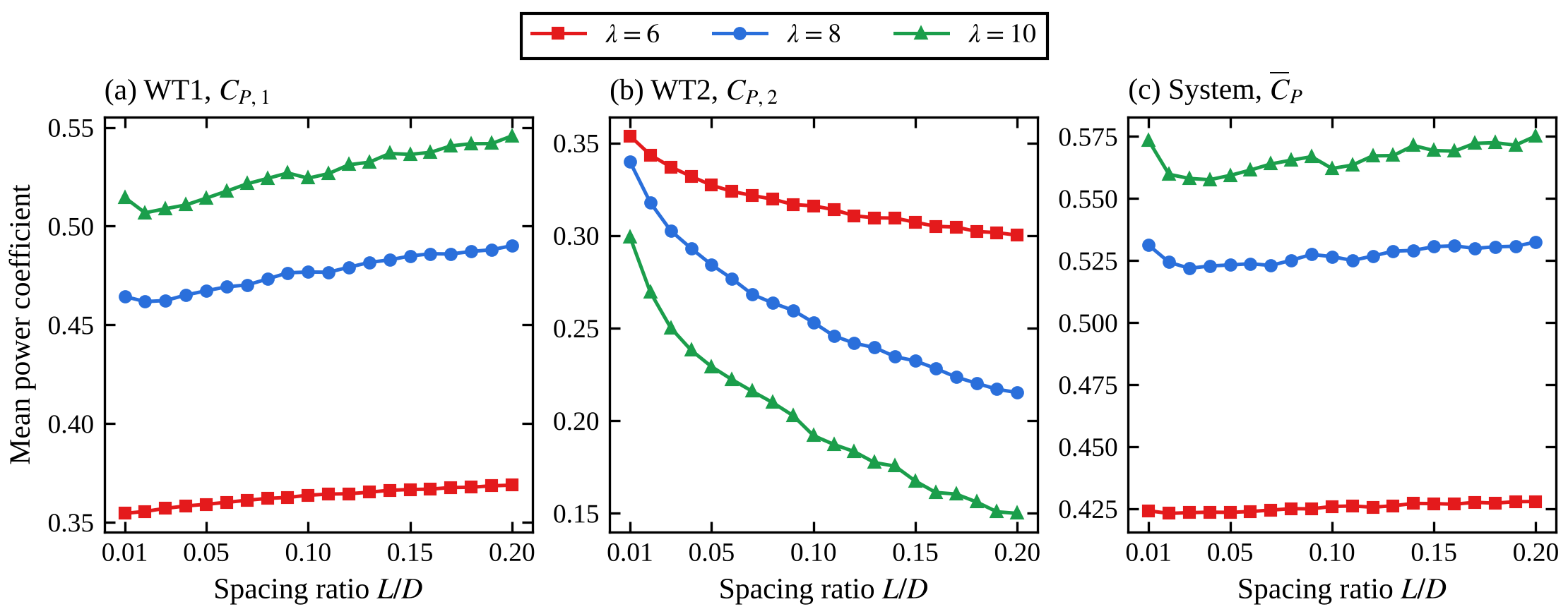}
\caption{Mean power coefficients from the MRUT grid simulations versus spacing ratio at $\lambda=6$, 8, and 10: (a) WT1, $C_{P,1}$; (b) WT2, $C_{P,2}$; (c) system, $\overline{C}_P$.}
\label{fig:rotorpower}
\end{figure*}

Figure~\ref{fig:rotorpower}
and Table~\ref{tab:rotorsplit} show how spacing redistributes power
between the rotors in the MRUT data. As $L/D$ increases from 0.01 to
0.20, the mean power of WT1 rises by 0.4--6.1\%, whereas that of WT2
falls by 0.9--61.8\%. Both changes generally become larger at higher
$\lambda$. Two mechanisms
are consistent with these trends. The downstream rotor decelerates the
flow upstream of its own plane, and this induction at the WT1 plane
weakens as the rotors move apart, so WT1 power rises with spacing.
Behind WT1, axial velocity decreases as the wake expands. Momentum
theory gives $U_\infty(1-a)$ at the rotor and $U_\infty(1-2a)$ in the
far wake, where $a$ is the axial induction factor \cite{Burton2011}.
WT2 therefore meets a slower inflow as spacing grows. WT1 delivers
73--96\% of the total power, so its gain largely offsets the WT2 loss.
At fixed $\lambda$, the system power coefficient $\overline{C}_P$ varies
by at most 3.9\% over the spacing range, and $\overline{C}_T$ by at most 1.7\%.
Opposite trends of
upstream- and downstream-rotor power with spacing were also reported for
a 77 m/56 m tandem pair \cite{Li2025} and in the model-scale tests described in Section~\ref{subsec:twin-rotor-model-and-the-vortex-method}. The preferred spacing thus results from a small
difference between two larger, opposing changes, which makes it
sensitive to $\lambda$ and to the selection rule.

\begin{table*}[!t]
\caption{Effect of spacing on the rotor power coefficients in the MRUT grid data, and spacing selected in the fixed-$\lambda$ search. Variations are $(\max-\min)/\max$ over $0.01\leq L/D\leq0.20$. Spacing: MRUT, grid point with the largest $\overline{C}_P$; GA, spacing selected by the surrogate-based search with the 0.5\% tolerance rule.}
\label{tab:rotorsplit}
\centering
\small\setlength{\tabcolsep}{4pt}\renewcommand{\arraystretch}{1.10}
\begin{tabular}{ccccccccccc}
\toprule
 & \multicolumn{3}{c}{\textbf{WT1, $C_{P,1}$}} & \multicolumn{3}{c}{\textbf{WT2, $C_{P,2}$}} & \multicolumn{2}{c}{\textbf{Variation over $L/D$}} & \multicolumn{2}{c}{\textbf{Spacing $L/D$}} \\
\cmidrule(lr){2-4}\cmidrule(lr){5-7}\cmidrule(lr){8-9}\cmidrule(lr){10-11}
$\lambda$ & $L/D=0.01$ & $L/D=0.20$ & \textbf{Change} & $L/D=0.01$ & $L/D=0.20$ & \textbf{Change} & $\overline{C}_P$ & $\overline{C}_T$ & \textbf{MRUT} & \textbf{GA} \\
\midrule
4 & 0.131 & 0.132 & $+0.4\%$ & 0.251 & 0.249 & $-0.9\%$ & 0.28\% & 0.14\% & 0.06 & 0.200 \\
5 & 0.243 & 0.248 & $+1.7\%$ & 0.340 & 0.328 & $-3.5\%$ & 0.65\% & 0.45\% & 0.11 & 0.200 \\
6 & 0.355 & 0.369 & $+4.0\%$ & 0.354 & 0.300 & $-15.2\%$ & 1.12\% & 0.85\% & 0.20 & 0.017 \\
7 & 0.422 & 0.442 & $+4.8\%$ & 0.348 & 0.248 & $-28.9\%$ & 1.51\% & 0.97\% & 0.19 & 0.016 \\
8 & 0.464 & 0.490 & $+5.5\%$ & 0.340 & 0.215 & $-36.7\%$ & 1.99\% & 1.09\% & 0.20 & 0.018 \\
9 & 0.498 & 0.525 & $+5.4\%$ & 0.326 & 0.179 & $-45.1\%$ & 2.31\% & 1.24\% & 0.01 & 0.019 \\
10 & 0.515 & 0.546 & $+6.1\%$ & 0.299 & 0.150 & $-49.9\%$ & 3.07\% & 1.65\% & 0.20 & 0.015 \\
11 & 0.518 & 0.544 & $+4.9\%$ & 0.278 & 0.106 & $-61.8\%$ & 3.86\% & 1.34\% & 0.01 & 0.018 \\
\bottomrule
\end{tabular}
\end{table*}

\begin{figure*}[!t]
\centering
\includegraphics[width=0.94\linewidth]{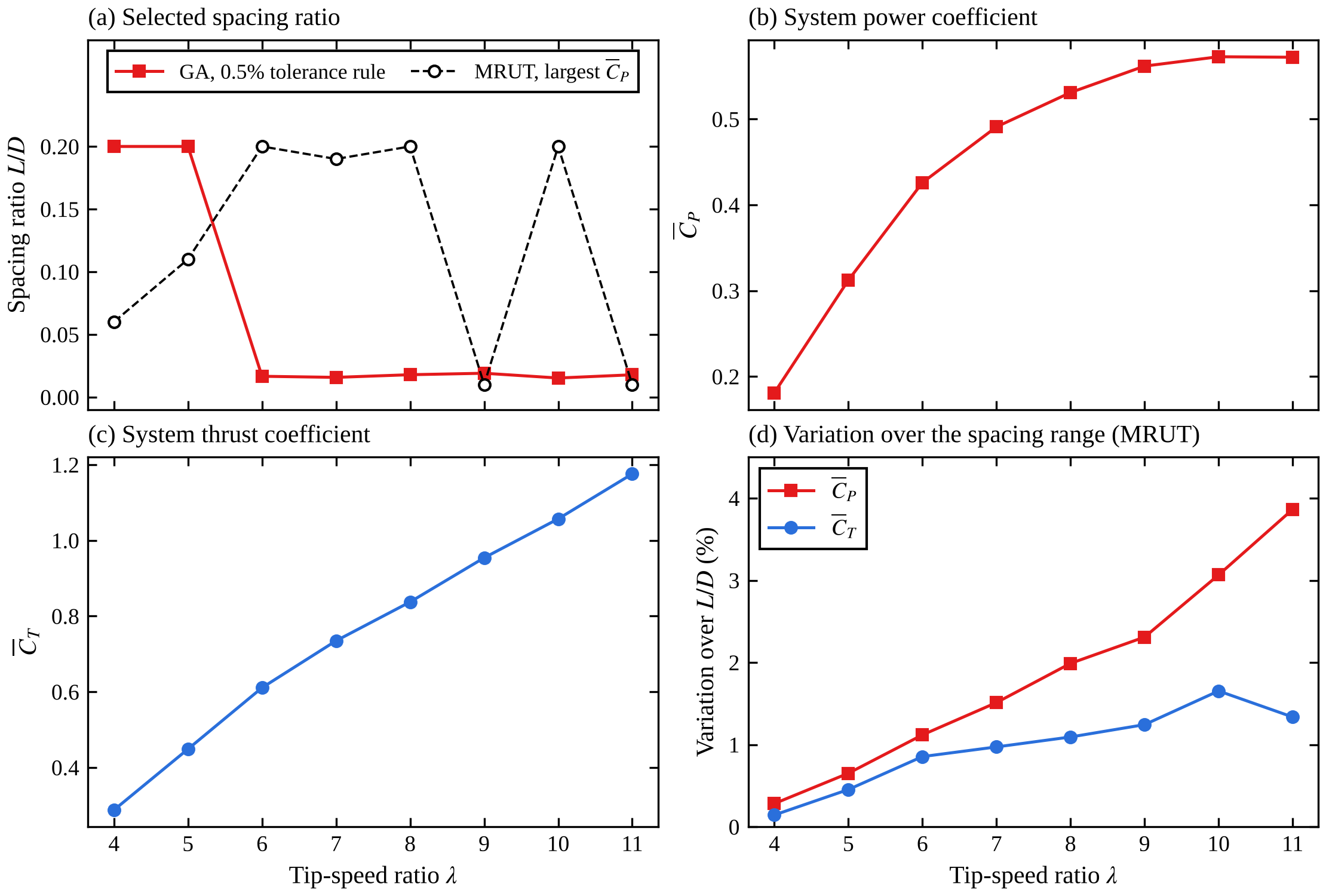}
\caption{Fixed-tip-speed-ratio search: (a) spacing ratio selected by the GA with the tolerance rule, and grid spacing with the largest $\overline{C}_P$ in the MRUT data; (b) system power coefficient $\overline{C}_P$ and (c) system thrust coefficient $\overline{C}_T$ of the selected configuration; (d) variation of $\overline{C}_P$ and $\overline{C}_T$ over $0.01\leq L/D\leq0.20$ in the MRUT data.}
\label{fig:9}
\end{figure*}

Figure~\ref{fig:9} shows the spacing
selected at each fixed $\lambda$ with the rule of
Section~\ref{subsec:twin-rotor-configuration-parameter-optimization-model-based-on-ga},
together with the corresponding $\overline{C}_P$ and $\overline{C}_T$. The
selected spacing is $L/D=0.20$ for $\lambda=4$ and 5 and lies between
0.015 and 0.019 for $\lambda=6$--11. The objectives are set almost
entirely by $\lambda$. $\overline{C}_P$ rises from 0.18 at $\lambda=4$
to 0.57 at $\lambda=10$ and then levels off.
$\overline{C}_T$ increases steadily from 0.29 to 1.18.
Figure~\ref{fig:9}d quantifies the weak dependence on spacing. The
cross-validated error of $\overline{C}_P$ is 0.23\% on average and 0.52\%
at the 90th percentile, so power differences between spacings are only
partly resolved. At $\lambda\geq6$, selection mainly follows the slightly
lower thrust at small spacing. In the MRUT grid data, the
spacing with the highest $\overline{C}_P$ is $L/D=0.20$ for $\lambda=6$, 8,
and 10, and $L/D=0.01$ for $\lambda=9$ and 11
(Table~\ref{tab:rotorsplit}). The two bounds differ in $\overline{C}_P$ by
less than 1\% except at $\lambda=11$ (1.5\%). With a shared tip-speed
ratio, $\lambda=4$ cannot be reached below rated wind speed, and the
other tip-speed ratios are reachable only in the wind-speed ranges of
Table~\ref{tab:rotor-speeds}.

\begin{figure}[!htb]
\centering
\includegraphics[width=\columnwidth]{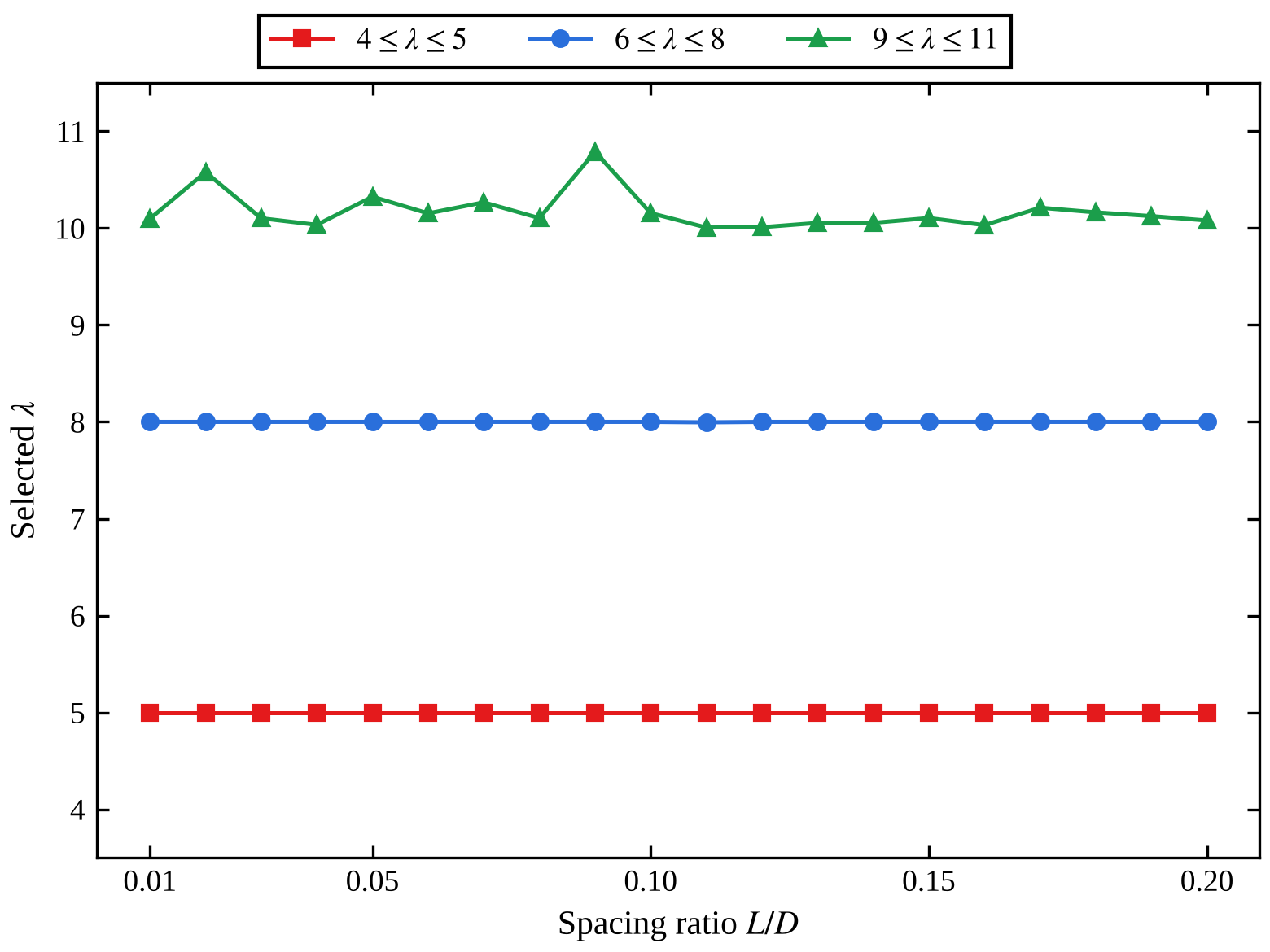}
\caption{Fixed-spacing search: tip-speed ratio selected within the low, medium, and high intervals at each spacing ratio.}
\label{fig:10}
\end{figure}

\begin{figure*}[!t]
\centering
\includegraphics[width=0.94\linewidth]{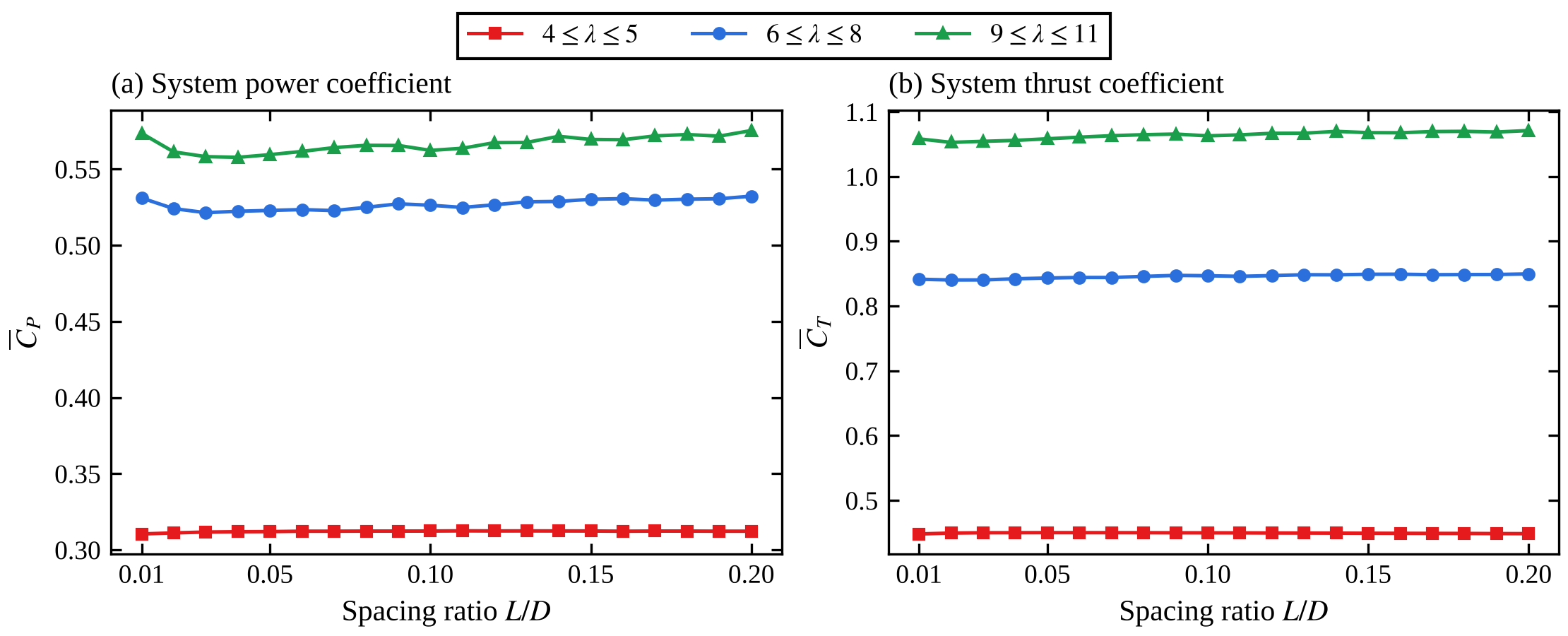}
\caption{Fixed-spacing search: (a) system power coefficient $\overline{C}_P$ and (b) system thrust coefficient $\overline{C}_T$ at the tip-speed ratio selected within each interval.}
\label{fig:11}
\end{figure*}

The complementary search fixes spacing and
varies $\lambda$ within low ($4\leq\lambda\leq5$), medium
($6\leq\lambda\leq8$), and high ($9\leq\lambda\leq11$) intervals.
Figure~\ref{fig:10} shows that the selected $\lambda$ lies at the upper
bound of the low and medium intervals for every spacing.
$\overline{C}_P$ increases with $\lambda$ there by much more than the
0.5\% tolerance. In the high interval, $\overline{C}_P$ peaks near $\lambda=10$
and changes little up to $\lambda=11$, whereas $\overline{C}_T$ keeps
increasing. The selected $\lambda$ therefore lies between 10.0 and 10.8.
At the selected $\lambda$, the low-interval $\overline{C}_P$ range is
0.310--0.313. The medium- and high-interval ranges are 0.521--0.532
and 0.558--0.575. The corresponding $\overline{C}_T$ ranges are
0.448--0.450, 0.840--0.850, and 1.05--1.07, respectively
(Figure~\ref{fig:11}). Moving from the
medium to the high interval at the same spacing raises $\overline{C}_P$
by about 8\% and $\overline{C}_T$ by about 26\%. In the medium and high intervals, $\overline{C}_P$ is highest at
$L/D=0.20$, almost as high at $L/D=0.01$, and lowest at
$L/D=0.03$--0.04. In the low interval, it varies by less than 0.7\%.
The selected $\lambda=5$ of the low interval is reachable only between
9.1 and 11.0 m/s (Table~\ref{tab:rotor-speeds}).

\begin{figure}[!htb]
\centering
\includegraphics[width=\columnwidth]{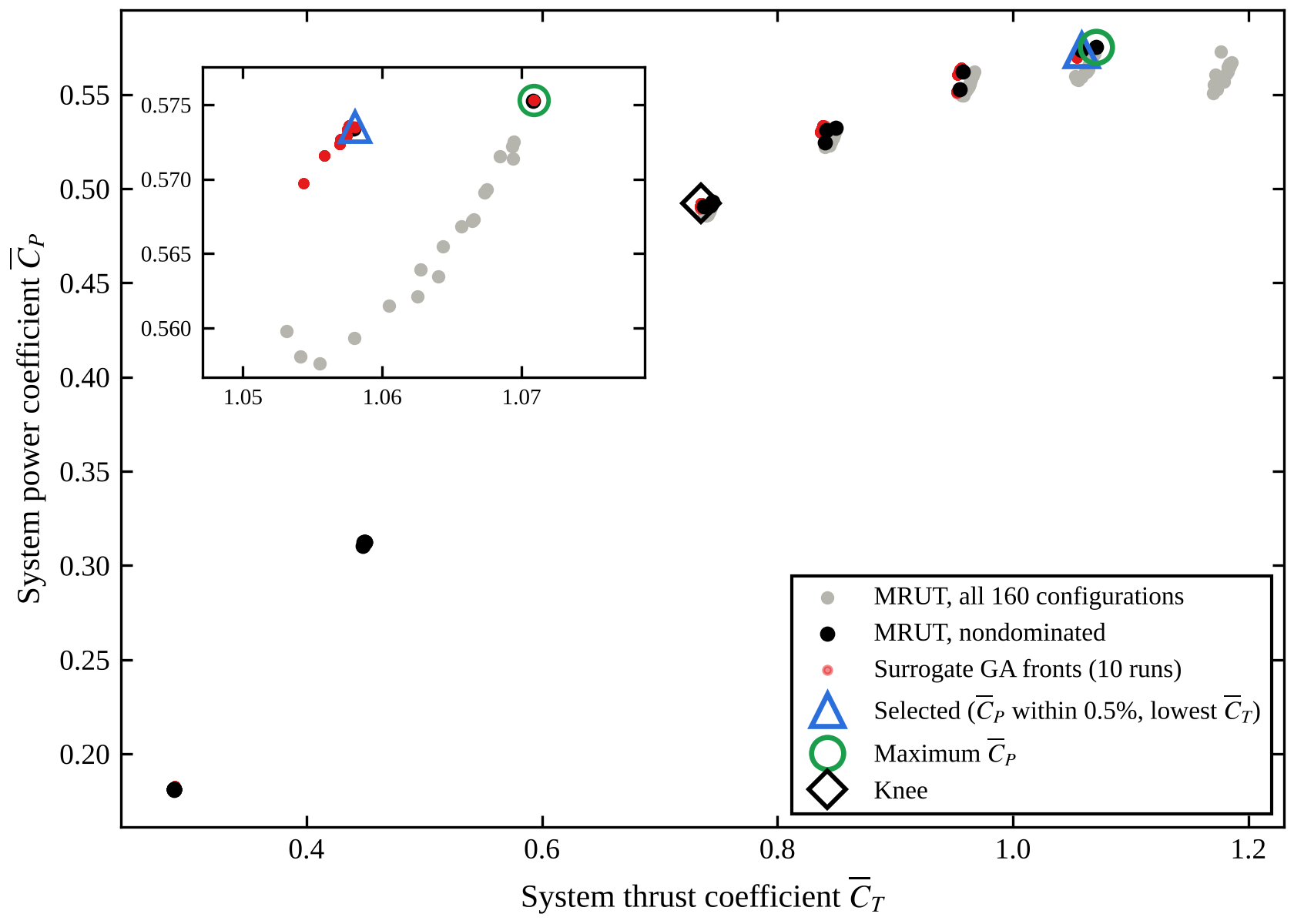}
\caption{Pareto front of the joint search in the system power and thrust coefficients. Grey and black dots show the 160 MRUT configurations and their nondominated subset. Red dots show the fronts of ten GA runs on the surrogate. The inset enlarges the high-power region.}
\label{fig:pareto}
\end{figure}

\begin{figure*}[!t]
\centering
\includegraphics[width=0.94\linewidth]{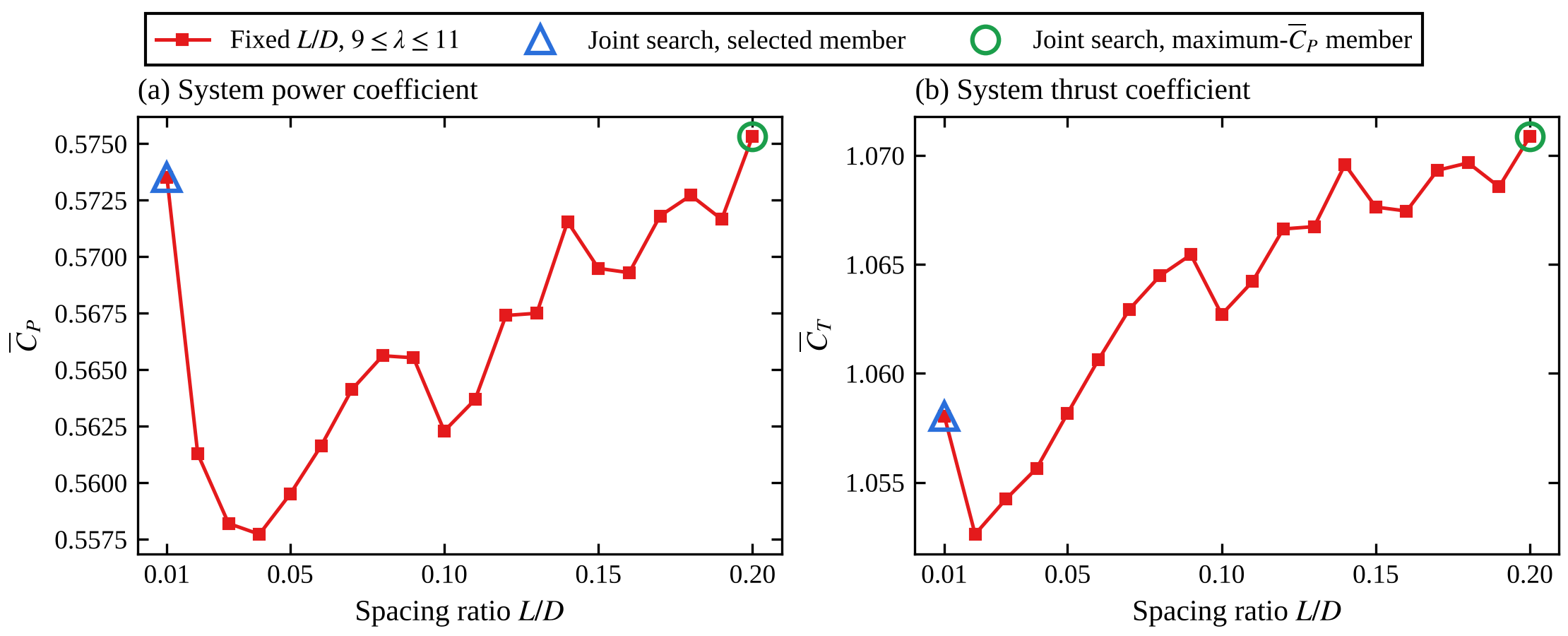}
\caption{Joint-search members compared with the high-interval fixed-spacing search: (a) system power coefficient $\overline{C}_P$; (b) system thrust coefficient $\overline{C}_T$. The selected member lies at $L/D=0.010$ and $\lambda=10.33$, the maximum-power member at $L/D=0.20$ and $\lambda=10.19$ in the run with random seed 0.}
\label{fig:12}
\end{figure*}

Figure~\ref{fig:pareto}
shows the fronts of ten joint searches with different random seeds,
together with all 160 MRUT configurations. The surrogate fronts follow
the nondominated MRUT configurations, and every run finds the same two
candidates at the high-power end of the front
(Figures~\ref{fig:pareto} and \ref{fig:12}). The maximum-power member lies
at $L/D=0.20$ and $\lambda=10.1$--10.4 in all runs, with
$\overline{C}_P=0.575$ and $\overline{C}_T=1.071$, fixed-pitch values summed
over both rotors and referred to the swept area of WT1
(Section~\ref{subsec:scope}). The selection rule
picks $L/D=0.010$--0.015 and $\lambda=10.0$--10.6 instead, where
$\overline{C}_P=0.573$ and $\overline{C}_T=1.057$--1.058. This gives
0.3--0.5\% less power for 1.2--1.3\% less thrust. The MRUT grid confirms that both are
nondominated. At $\lambda=10$, $\overline{C}_P=0.573$ and
$\overline{C}_T=1.058$ at $L/D=0.01$, and $\overline{C}_P=0.575$ and
$\overline{C}_T=1.071$ at $L/D=0.20$. WT2 contributes 0.059 and 0.030 to
$\overline{C}_P$, that is, 10\% and 5\% of the system power,
respectively
(Table~\ref{tab:rotorsplit}). The two candidates have similar mean power but differ in mean and
peak thrust. With $D$ equal to the
WT1 diameter of 284 m, $L/D=0.01$ leaves an axial distance of only
2.8 m between the rotor centres. Clearance for blade deflection and
nacelle hardware would also have to be checked. The knee is the member
farthest from the line joining the front's extreme members in normalized
coordinates. It lies at $\lambda\approx7.8$ and $L/D\approx0.013$, with
$\overline{C}_P=0.492$ and $\overline{C}_T=0.735$. It provides 86\% of
the maximum power for 69\% of the corresponding thrust. Within the
operating envelope, $\lambda=10$ requires $4.6\leq U_\infty\leq8.0$ m/s
(Table~\ref{tab:rotor-speeds}). At 8.0 m/s, the configuration at
$L/D=0.20$ corresponds to 10.8 MW and 2.32 MN for WT1 and 0.58 MW and
0.33 MN for WT2. The WT1 thrust is below the rated thrust of 2.79 MN of
the IEA Wind 22 MW rotor \cite{ref31}. The knee at $\lambda\approx7.8$ is reachable
between 5.8 and 10.2 m/s.

The maximum-power member lies at the upper spacing bound, and in
the MRUT data at $\lambda=10$, $\overline{C}_P$ still increases from
$L/D=0.19$ to 0.20 (by 0.7\%). This optimum is therefore boundary-limited.
Both the WT1 gain and the WT2 loss continue at the bound
(Figure~\ref{fig:rotorpower}). Their balance beyond $L/D=0.20$ is
unknown, so the present data cannot establish whether $\overline{C}_P$
continues to rise. The peak load on the shared support is measured here by the
peak of the instantaneous system thrust coefficient,
$C_{T,\mathrm{peak}}=\max_k\,(T_{1,k}+T_{2,k})/(\tfrac{\pi}{8}\rho
U_\infty^2D_1^2)$. Replacing $\overline{C}_T$ by $C_{T,\mathrm{peak}}$
as the second objective changes the number of
nondominated MRUT configurations from 39 to 38 and removes the
small-spacing candidate. At $\lambda=10$, $C_{T,\mathrm{peak}}$ is 1.195
at $L/D=0.01$ and 1.186 at $L/D=0.20$.
The latter configuration has higher mean power and lower peak thrust.
The margin is 0.8\%. Individual rotor peaks differ more because WT2
thrust fluctuates strongly at the smallest spacing. At $\lambda=10$,
its standard deviation is 22\% of its mean at $L/D=0.01$ and 11\% at
$L/D=0.20$. The sum of the two rotor maxima, normalized in the same
way, is 1.26 and 1.20, respectively. The first stage reproduces
$C_{T,\mathrm{peak}}$ of unseen
configurations with a cross-validated error of 0.42\% on average and
0.85\% at the 90th percentile (evaluated on every fourth time node).
This error is comparable to the 0.8\% margin, so the ranking under the
peak criterion rests on the MRUT data rather than on the surrogate.

\subsection{Computational cost}\label{subsec:cost}

For the validation case of Table~\ref{tab:validation}, MRUT used
30.8 h of total CPU time on an eight-core AMD 9700X processor.
The RANS solution used 11,616 h on two 48-core AMD EPYC processors.
Surrogate timings were measured on a two-core Intel Xeon virtual machine
with a 2.1 GHz processor and 8 GB memory. Training the 4236 per-node
XGBoost models for one rotor and response took 66 s. Predicting a new
configuration's 600 s history took 3.4 s.
Training the temporal network of Section~\ref{subsec:benchmark} on one
history took 28 s.
Tabulating the first-stage objectives on a grid of 134,000
configurations took 6 min, after which a GA run with 1050 evaluations
took less than 1 s. The vortex simulations therefore dominate the total
cost, and the surrogate pays off when many configurations or repeated
searches are evaluated.

\subsection{Scope and limitations}\label{subsec:scope}

The results describe a simulation-based aerodynamic design study.
The simulations use fixed blade pitch and prescribed rotor speed
without the pitch and torque controllers of the reference turbines. A
real turbine can run at a given tip-speed ratio only within its
rotor-speed range and, with fixed pitch, only below rated wind speed.
The coefficients apply only inside the envelope of
Figure~\ref{fig:envelope} and Table~\ref{tab:rotor-speeds}, which
excludes $\lambda=4$. The highest system power coefficient, 0.575 at
$\lambda=10$, is below the Betz limit of $16/27\approx0.593$ for a single
actuator disc \cite{Burton2011} and below $16/25=0.64$ for two discs in
tandem \cite{Newman1986}. At $\lambda=9$--11, however, the WT1 power
coefficient (0.53--0.55 at $L/D=0.20$) is up to 9\% higher than the value
of about 0.5 at which the IEA Wind 22 MW rotor operates below rated wind
speed \cite{ref31}. The WT1 thrust coefficient is 0.94 at $\lambda=10$,
against about 0.8 for the reference rotor below rated wind speed
\cite{ref31}, and reaches 1.0 at $\lambda=11$, a heavily loaded state in
which the lifting-line model is less reliable. The coefficients at
$\lambda\geq10$ are best read as upper estimates.
MRUT represents viscous effects
through vortex-core growth and does not model the nacelle, the tower,
or blade flexibility. Its mean rotor loads agree with RANS at
$L/D=0.1$ under uniform inflow, but its near-wake profiles and load histories are smoother
than the CFD solution. Model-scale tests showed an overprediction of
the downstream-rotor thrust (Section~\ref{subsec:twin-rotor-model-and-the-vortex-method}).
The WT2 results are therefore more uncertain than those of WT1. All
simulations use one inflow realization at 11.4 m/s and 10\% turbulence
intensity. Other wind speeds and turbulence levels change the induction
of both rotors and lie outside the present data. The results also apply
to the present rotor pair. WT2 (126 m) spans 44\% of the WT1 diameter
(284 m) \cite{ref31,ref32} and operates in the inner part of the WT1
wake. Pairs with another diameter ratio, or with the smaller rotor
upstream, sample different parts of the wake
\cite{Rosenberg2014,Salah2024,Li2025} and would require new training data.

The first stage interpolates within the sampled
grid but does not extrapolate beyond it (Table~\ref{tab:cv}). Because all
histories share one wind realization, the second-stage tests measure how
well the network reproduces the response to that realization. Forecasting
under an independent realization has not been tested. The architecture
comparison of Section~\ref{subsec:benchmark} uses horizons of up to
1.42 s with simulated inputs. The free-running 600 s predictions of
Section~\ref{subsec:multi-parameter-power-prediction-for-twin-rotor-wind-turbines}
were made with the CNN-Attention-LSTM only. The WT2 branch of the
second stage requires the vortex-method descriptors
$\mathbf{F}_{VM}$, so its application to a new configuration includes a
vortex-method computation of these quantities.

The objectives use the time-mean
thrust, and the peak thrust serves only to separate candidates of
nearly equal power. Fatigue loading, tower deflection, and platform
loads are outside the present constraints. On a floating support,
wave-driven platform motion can raise rotor thrust fluctuations
substantially \cite{Yan2026}. A design study for a specific turbine would
need data spanning operating wind speeds and turbulence intensities,
with several wind realizations and a wider spacing range. It would
also need an aeroelastic model of the shared support with strength and
fatigue constraints. Selected configurations would require direct
vortex or CFD simulation.

\FloatBarrier

\section{Conclusions}\label{sec:conclusion}

This study combined vortex-method simulations of an unequal tandem
rotor pair, a configuration-conditioned surrogate of the response
histories, and a genetic algorithm. The main findings are as follows.

Spacing changes the distribution of power between the rotors more
strongly than their combined power. From $L/D=0.01$ to 0.20, WT1 power
rises by up to 6.1\% and WT2 power falls by up to 62\%. At fixed
$\lambda$, the system power coefficient varies by at most 3.9\% and the
system thrust coefficient by at most 1.7\%. Tip-speed ratio sets the power level and thrust.

The first stage predicts the power histories of unseen grid
configurations with a median $R^2$ of 0.995 for WT1 and 0.958 for WT2.
The WT2 errors concentrate at $L/D\leq0.02$ and high $\lambda$, and the
accuracy collapses outside the sampled range of $\lambda$. On the regular
grid, linear interpolation is about as accurate as XGBoost. It gives
higher median $R^2$ and lower WT2 errors, whereas XGBoost has lower
mean WT1 error. In the temporal stage, persistence matches
CNN-Attention-LSTM one step ahead for WT1 but is less accurate than the
networks for WT2. Ten steps (1.42 s) ahead, all tested networks clearly
outperform persistence. The convolutional layer provides most of the
gain over a vanilla LSTM.

Under the system power and thrust coefficient objectives, the
high-power end of the Pareto front lies near $\lambda=10$. At
$L/D=0.20$, the fixed-pitch system coefficients referred to the swept
area of WT1 are $\overline{C}_P=0.575$ and $\overline{C}_T=1.07$. At $L/D\approx0.01$, 0.3--0.5\% less power comes
with 1.2--1.3\% less mean thrust but a slightly higher peak thrust, so
a peak-load criterion favours $L/D=0.20$. The knee near $\lambda\approx7.8$ gives 86\% of
the maximum power for 69\% of the thrust. With a shared tip-speed ratio, the two
reference rotors can operate at $\lambda=10$ only between 4.6 and
8.0 m/s and at the knee between 5.8 and 10.2 m/s. They cannot reach
$\lambda=4$ below rated wind speed. These results identify
aerodynamically favourable regions. Their structural adequacy requires
peak-load, fatigue, and platform-load analysis with an aeroelastic model
of the shared support.

\begin{acknowledgments}
This work was performed within the Strategic Research Plan of the Centre for Marine Technology and Ocean Engineering (CENTEC). Yinong Tian was supported by the Postdoctoral Project of the China Scholarship Council (Grant No.\ 202506680040).
\end{acknowledgments}

\section*{Author Declarations}

\subsection*{Conflict of Interest}

The authors declare that they have no known competing financial interests or personal relationships that could have appeared to influence the work reported in this paper.

\subsection*{Author Contributions}

\textbf{Yinong Tian:} Conceptualization; Methodology; Software; Validation; Writing -- original draft. \textbf{Shan Wang:} Conceptualization; Resources.

\subsection*{Use of generative AI}

During manuscript preparation, the authors used OpenAI Codex (accessed September 2026) to assist with English-language editing and LaTeX formatting. The authors are responsible for reviewing and editing the output and take full responsibility for the content of this publication.

\section*{Data Availability}

The code and data supporting this study are publicly available at \url{https://github.com/robytian/ML-tandem-rotor}.

\bibliography{references}

\end{document}